\documentclass[journal,12pt,onecolumn,draftclsnofoot,]{IEEEtran}
\usepackage{xcolor,soul,framed}  
\usepackage[noadjust]{cite}
\usepackage{filecontents}

\usepackage[pdftex]{graphicx}
\graphicspath{{../pdf/}{../jpeg/}}
\DeclareGraphicsExtensions{.pdf,.jpeg,.png}

\usepackage{amssymb,mathtools}
\usepackage{array}
\usepackage{mdwmath}

\usepackage{eqparbox}
\usepackage{url}

\usepackage{enumitem} 
\usepackage{tabularx} 
\usepackage{multirow}  
\usepackage{booktabs} 
\usepackage{makecell} 
\usepackage{aligned-overset}

\usepackage{url}
\usepackage{multicol} 
\usepackage{balance}
\usepackage{bbm}
\usepackage{amsmath}
\usepackage{amsthm} 
\usepackage[caption=false, font=footnotesize]{subfig}
\usepackage[ruled,vlined,linesnumbered]{algorithm2e}
\SetKwRepeat{Do}{do}{while}
\SetKwBlock{Begin}{begin}{end}

\graphicspath{{./Figures/}}
\newtheorem{Remark}{Remark}

\newtheorem{Definition}{Definition}
 \newtheorem{Theorem}{Theorem}

\renewcommand{\vec}[1]{\mathbf{#1}}

\newcommand{\DIRS}{\mathrm{DR}}
\newcommand{\UIRS}{\mathrm{UR}}
\newcommand{\uu}{\mathrm{UD}}
\newcommand{\du}{\mathrm{DD}}

\newcommand{\AP}{\mathrm{AP}}

\newcommand{\pl}{\ell}

\usepackage[left=.5in,                               right=.5in,                     top=1in,  bottom=1in]{geometry}

\begin{document}
   \title{Joint Devices and IRSs Association for Terahertz Communications in Industrial IoT Networks}
 \author{         {Muddasir Rahim, Georges~Kaddoum,~\IEEEmembership{Senior~Member,~IEEE}, and Tri~Nhu~Do}
\thanks{M.~Rahim,~G.~Kaddoum,~and~T.~N.~Do are with the Department of Electrical Engineering, the \'{E}cole de Technologie Sup\'{e}rieure (\'{E}TS), Universit\'{e} du Qu\'{e}bec, Montr\'{e}al, QC H3C 1K3, Canada (emails: muddasir.rahim.1@ens.etsmtl.ca, georges.kaddoum@etsmtl.ca, tri-nhu.do@etsmtl.ca)}} \maketitle
\begin{abstract}
The Industrial Internet of Things (IIoT) enables industries to build large interconnected systems utilizing various technologies that require high data rates. Terahertz (THz) communication is envisioned as a candidate technology for achieving data rates of several terabits-per-second (Tbps). Despite this, establishing a reliable communication link at THz frequencies remains a challenge due to high pathloss and molecular absorption. To overcome these limitations, this paper proposes using intelligent reconfigurable surfaces (IRSs) with THz communications to enable future smart factories for the IIoT. In this paper, we formulate the power allocation and joint IIoT device and IRS association (JIIA) problem, which is a mixed-integer nonlinear programming (MINLP) problem. {Furthermore, the JIIA problem aims to maximize the sum rate with imperfect channel state information (CSI).} To address this non-deterministic polynomial-time hard (NP-hard) problem, we decompose the problem into multiple sub-problems, which we solve iteratively. Specifically, we propose a Gale-Shapley algorithm-based JIIA solution to obtain stable matching between uplink and downlink IRSs. {We validate the proposed solution by comparing the Gale-Shapley-based JIIA algorithm with exhaustive search (ES), greedy search (GS), and random association (RA) with imperfect CSI.} The complexity analysis shows that our algorithm is more efficient than the ES. 
\end{abstract}


\section{INTRODUCTION}
\IEEEPARstart{F}{ollowing} the rapid development of smart manufacturing and industrial automation, the aim is to increase productivity and efficiency by integrating the physical world with the internet through the Industrial Internet of Things (IIoT)~\cite{sisinni2018industrial}. Thousands of industrial devices are now connected to the internet through the IIoT, enabling them to transmit and gather data that can be used by business processes and services. However, the increasing demand for cutting-edge services in innovative industries, like augmented reality (AR)/virtual reality (VR) maintenance, automatic guided vehicles, and holographic control display systems, will pose substantial communication challenges for industrial wireless communications~\cite{letaief2019roadmap}. The large-scale IIoT networks, being extremely efficient, are intended to provide extremely high-speed communication of up to terabits-per-second (Tbps) with low latency, ultra-high reliability, and minimal power consumption~\cite{zeb2022industrial, rubab2022interference,mao2021energy}. 

A promising candidate frequency band for wideband communication is terahertz (THz), which can deliver ultra-high transmission rates, i.e., several Tbps~\cite{chen2019survey,wan2021terahertz}. However, despite the many advantages of the THz band, establishing a reliable transmission link at THz frequencies remains a challenge in industrial wireless communications. There are several factors that contribute to the challenges of establishing a reliable THz transmission link in industrial wireless communications, including metallic structures, arbitrary movement of IIoT devices, such as vehicles and robots, factory dimensions, and electromagnetic interference. Moreover, beam alignment is extremely challenging due to the narrow beamwidth in THz bands~\cite{sarieddeen2019terahertz, DoTCOMM2021,han2016distance}.

Intelligent reconfigurable surfaces (IRSs) are gaining increasing attention from academic as well as industrial communities as a promising technology for providing smart and reconfigurable industrial wireless communication environments~\cite{wu2021intelligent}. The IRS consists of a planar metasurface with multiple passive reflecting elements. These elements can be digitally controlled in order to alter the wireless channel between transmitters and receivers by varying the amplitude and/or phase shift (PS) of the incident signal. This means IRSs have the potential to reshape the industrial wireless propagation environment in favor of signal transmission, which is fundamentally different from existing techniques~\cite{di2020reconfigurable}. Additionally, IRSs have shown potential in managing dynamic networks, such as dealing with the movement of nodes, changes in network topology, and network topology reconstruction. IRSs are used to enhance the coverage area of wireless networks by reflecting and redirecting the signals to areas that are difficult to reach~\cite{wu2019intelligent}. This can be especially useful in urban areas or indoor environments where signals can be obstructed or weakened by obstacles. IRSs can be used in device-to-device (D2D) communication to enhance network throughput by creating temporary links between devices~\cite{shah2022statistical}. Furthermore, IRSs can help to mitigate interference caused by multi-path propagation or co-channel interference. By creating virtual propagation paths, IRSs can separate signals from different sources, which can help to improve the signal-to-noise ratio (SNR) and reduce interference~\cite{elmossallamy2020reconfigurable}.

Moreover, IRSs also enjoy additional practical advantages, such as having a low profile, lightweight, and conformal geometry, which make them suitable for wide-scale deployment across industrial wireless networks. Furthermore, an IRS-assisted network can increase the efficiency of the IIoT network in terms of data rate, connectivity, and coverage. Considering these facts, we integrated IRSs into the IIoT networks to facilitate both uplink and downlink communications. However, optimizing the association time is crucial for achieving optimal network performance. This is particularly important when the network topology is subject to frequent changes due to the movement of nodes. Furthermore, the association time depends on various network conditions, including node density and mobility. Therefore, it is important to implement strategies that can reduce the association time to improve the network performance.
\subsection{{Related Works}} 
Recent studies have explored IRSs, potential for improving industrial wireless communication. Specifically, research works, such as~\cite{dhok2021non,hashemi2021average, li2022exploiting, cheng2022robust, firyaguna2022towards,schellmann2022capacity}, have studied the energy harvesting, average rate, and reliability analysis of IRS-assisted industrial networks. For example, S. Dhok \textit{et al.} integrated IRSs in an industrial environment to assist the communication between the data center (DC) and the server machine (SM)~\cite{dhok2021non}. This work aimed to maximize the system reliability and energy efficiency by optimizing the hybrid power-time splitting and channel power-based scheduling of destinations. In~\cite{hashemi2021average}, the authors investigated the average achievable rate and error probability of IRS-assisted systems in a factory automation scenario under the finite blocklength (FBL) regime. Work~\cite{hashemi2021average} also derived analytical expressions for both the average achievable rate and error probability. {X. Li \textit{et al.} considered an IRS-based wireless-powered non-orthogonal multiple access (NOMA) IoT network to maximize network sum throughput~\cite{li2022exploiting}. The idea behind this work was to use an alternative optimization (AO) technique to optimize the PS matrices for wireless energy transfer (WET) and wireless information transfer (WIT).}

J. Cheng \textit{et al.} conducted a study demonstrating the reliability between the access point (AP) and actuators of multiuser multi-input-single-output (MISO) in the IRS-assisted IIoT network~\cite{cheng2022robust}. The idea of this work was to jointly optimize the active and passive beamforming, which aims to maximize the number of actuators with successful decoding for both perfect and imperfect channel state information (CSI). In~\cite{firyaguna2022towards}, the authors demonstrated the potential features and advantages of IRSs in a smart factory environment to support the evolution towards Industry $5.0$. Additionally, this work elaborated on the challenges and potential opportunities for research into future IRS-assisted wireless factory automation. Finally, M. Schellmann~\textit{et al.} considered an IRS-aided IIoT communication for a factory environment to maximize the system capacity~\cite{schellmann2022capacity}. The idea of this work was to derive path-loss models for indoor factory scenarios with multiple IRSs to assist industrial wireless communications to maximize the network capacity. 
\subsection{{Contributions}}
The aforementioned studies focused on investigating the performance of IRS-assisted networks by optimizing the beamforming matrix at the IRSs. However, the joint IIoT device and IRS association (JIIA), a critical factor affecting industrial wireless communications in THz bands, is largely ignored in the existing literature. Industrial application demand for high data rates with good coverage in mobile IIoT device scenarios. Thus, in this work, the IRS is used to assist the uplink and downlink THz band communication between the IIoT devices and the AP in an industrial environment. In this context, we formulate an optimization problem to maximize the sum rate in industrial IoT networks. This makes it a promising solution for improving the performance of industrial IoT networks, which are increasingly being used in various industries. The key contributions of this paper are:
\begin{itemize}
    \item {To the best of the authors' knowledge, this paper is the first work to study the power allocation and the JIIA problem in IRS-assisted mobile IIoT networks operating at THz bands with imperfect CSI}. To maximize the sum rate, we formulate the JIIA problem as a mixed-integer nonlinear programming (MINLP) problem, which is challenging to solve. Additionally, the IIoT devices are mobile, which means that the channel condition between IIoT devices and IRSs may frequently change, requiring a robust algorithm for resource allocation algorithm including power allocation and IRSs association problem. 
    \item A new framework to tackle the formulated optimization problem. We first decompose the MINLP problem into multiple sub-problems and then obtain the optimal PS configuration with a diffuse reflection model. The optimal decoding and beamforming vectors are achieved using the minimum-mean-square-error (MMSE) method. Furthermore, we propose an optimal power allocation algorithm using a water-filling scheme and present the analytical expression for the optimal power allocation, which facilitates faster algorithm execution.
    \item We then propose the Gale-Shapley-based robust JIIA algorithm, which obtains a near-optimal solution to the original problem and significantly reduces the computational complexity. It is proved that the proposed JIIA algorithm converges to a stable matching and terminates after a finite number of iterations in a mobile IIoT scenario.
    \item Numerical results demonstrate that the proposed JIIA can significantly improve the sum rate compared to the greedy search (GS) and random association (RA) approaches. Additionally, the sum rate of our proposed method is comparable to that of the exhaustive search (ES) method, without considering the association overhead. This demonstrates the optimality of the proposed JIIA algorithm. However, when taking into account the association overhead, the sum rate of the JIIA scheme outperforms the GS, RA, and ES schemes because it minimizes the negative impact of association overhead.
\end{itemize}
\section{System Model} \label{model}
\begin{figure}[!t]
     \centering
     \includegraphics[width=0.5\linewidth]{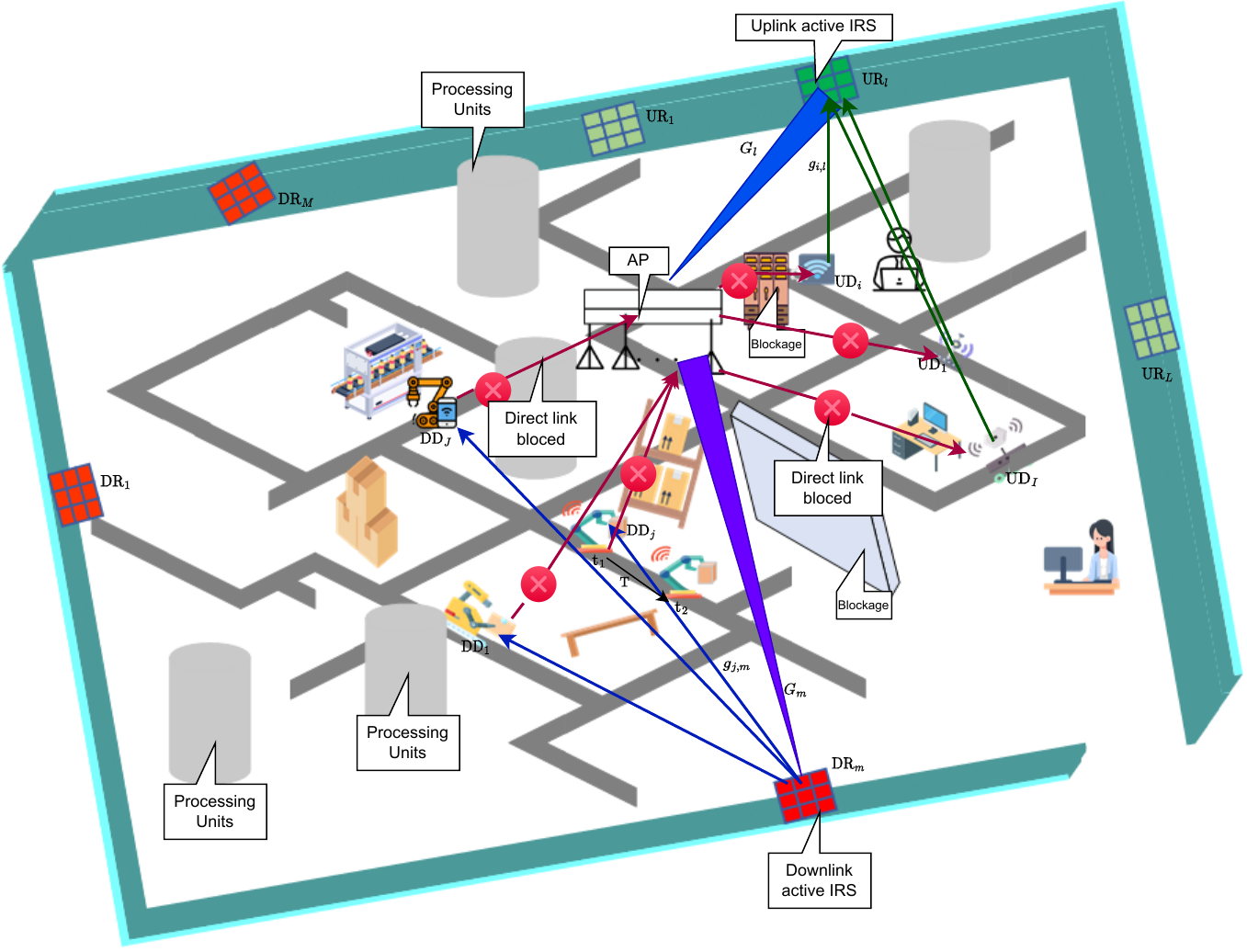}
     \caption{Deployment of IRS-assisted THz communications for autonomous robots in an industrial environment.
     }
     \label{m1}
\end{figure}
 We consider IRS-assisted autonomous robots in an IIoT network over THz bands. As shown in Fig.~\ref{m1},  the access point (AP) is equipped with $K$ antennas for uplink and downlink communications. {Moreover, we assume that the AP operates in half-duplex (HD) mode and can thus either receive data from the uplink IIoT device (UD) or transmit data to the downlink IIoT device (DD) at any one moment.} We deployed $I$ uplink IIoT devices and $J$ downlink IIoT devices, with a single antenna each. In addition, $L$ uplink IRSs and $M$ downlink IRSs are used to assist the uplink and downlink communications, respectively, where each IRS is equipped with $N$ reflecting elements. It is assumed that the direct links from IIoT devices and the AP are unavailable due to unfavorable propagation conditions\footnote{{This assumption is reasonable when the direct link is severely impeded by obstacles or human bodies in the environment~\cite{wan2021terahertz,fang2022optimum}.}}. Therefore, the uplink and downlink communication takes place via IRS deployed in the network. The area of each IRS element is $A = N_x \times N_y$, where $N_x$ and $N_y$ represent the length of the horizontal and vertical sides, respectively. In the three-dimensional (3D) Cartesian coordinate system, the locations of the $i$-th uplink device ($\uu_i$) and the $j$-th downlink device ($\du_j$) at time slot $t$ are denoted by $\vec c_{i}(t) = (x_i(t), y_i(t), z_i)$ and $\vec c_j(t) = (x_j(t), y_j(t), z_j)$, respectively. Moreover, the locations of the $l$-th uplink IRS ($\UIRS_l$), the $m$-th downlink IRS ($\DIRS_m$), and the AP are defined as $\vec c_l = (x_l, y_l, z_l)$, $\vec c_m = (x_m, y_m, z_m)$, and $\vec c_\AP = (x_\AP, y_\AP, z_\AP)$, respectively. Let $d_{i,l,n}(t)$ denotes the distance from $\uu_i$ to the $n$-th element of the $l$-th uplink IRS ($\UIRS_{l,n}$) at time slot $t$, and $d_{l,n}$ denotes the distance from $\UIRS_{l,n}$ to the AP. For THz communications, the IRS elements are densely deployed, and the relative distance between the IRS elements is very short. Therefore, we can ignore the distance between the IRS elements. Thus, the distance $d_{i,l}(t)$ can be computed from the location of $\UIRS_l$ to $\uu_i(t)$, which can be expressed as
$    d_{i,l}(t) = \Vert \vec{c}_i(t) - \vec{c}_l \Vert
    = \sqrt{(x_i(t)-x_l)^2 + (y_i(t)-y_l)^2 + (z_i-z_l)^2}.
$
Similarly, let $d_{m,n}$ denotes the distance from the AP to the $n$-th element of the $m$-th downlink IRS ($\DIRS_{m,n}$), and $d_{j,m,n}(t)$ denotes the distance from $\DIRS_{m,n}$ to the $j$-th downlink device $\du_j$ at time slot $t$.

\subsection{{Uplink Modeling of the Considered IRS-THz Network}}
We assume that the direct links between the IIoT devices and the AP are blocked by an obstacle and thus focus on the propagation model for cascaded links. {Furthermore, we assume that the CSI available at the IIoT devices and the AP is imperfect\footnote{{Channel estimation is much more difficult in IRS-assisted wireless communication than in traditional wireless networks. Furthermore, it is more challenging at the IRS than the AP because the IRS has passive reflecting elements that cannot process the pilot signals sent to and from the IIoT devices / AP~\cite{pan2021reconfigurable, shah2022statistical}. However, channel reciprocity holds for the IIoT device-to-IRS and IRS-to-AP channels~\cite{tang2021channel}. We therefore use the IRSs' estimated CSI at the IIoT devices / AP.}}. Let $\widehat{\vec{g}}_{i,l} \in \mathbb{C}^{N \times 1}$ denote the estimated channel vector from $\uu_i$-to-$\UIRS_l$. In this context, $[\widehat{\vec{g}}_{i,l}]_n = \widehat{g}_{i,l,n}$ is the individual channel from $\uu_i(t)$-to-$\UIRS_{l,n}$, which can be written as 
$\widehat{g}_{i,l,n}(t) = \sqrt{\pl_{i,l,n}(t)} e^{ - j \omega d_{i,l,n}(t)},$ where $\omega = \frac{2 \pi}{\lambda}$ [m${}^{-1}$] is the wavenumber and $\pl_{i,l,n}$  represents the pathloss of the $\uu_i$-to-$\UIRS_{l,n}$ link.  Furthermore, it is important to study the impact of channel estimation errors (CEEs) to improve optimization design. Then, we define channel with CEEs as the composition of the estimated channel and its corresponding CEEs, which is given by
\begin{align}\label{eq:2}
{g}_{i,l,n}(t) = \widehat{g}_{i,l,n}(t)+\widetilde{g}_{i,l,n}(t),
\end{align}
where $\widetilde{g}_{i,l,n}(t)$ denotes the CEEs, which are uncorrelated with the estimated channel ${g}_{i,l,n}(t)$ and the entries of $\widetilde{\vec{g}}$ are independent and identically distributed (IID) complex Gaussian with zero mean and variance $\sigma_{\widetilde{\vec{g}}}^2$. The correlation between the estimated and actual channel CSI, which is assumed to be the same for all gains, is given by
\begin{align}\label{cor1}
\rho_{\vec{g}}
    =   \frac{\mathbb{E}\{ [\vec{g}]_{i} [\widehat{\vec{g}}]_{i}^\ast \}}{\sqrt{\mathbb{E}\{ |[\vec{g}]_{i}|^2 \} \mathbb{E}\{ |[\widehat{\vec{g}}]_{i}|^2 \} }}
    =   \frac{1}{\sqrt{1+\sigma_{\vec{g}}^2}},
\end{align}
Let $[\widehat{\vec{g}}_{l,k}]_n =  \widehat{g}_{l,n,k}$ be an individual estimated channel from $\UIRS_{l,n}$ to antenna $k$ of the AP, where $\vec{g}_{l,k} \in \mathbb{C}^{N \times 1}$ is the channel vector from $\UIRS_l$ to antenna $k$ of the AP, which can be written as $\widehat{g}_{l,n,k} = \sqrt{\pl_{l,n,k}} e^{ - j \omega d_{l,n,k}},$ where $\pl_{l,n,k}$  represents the pathloss of the $\UIRS_{l,n}$-to-antenna $k$ link. Moreover, with $\widetilde{g}_{l,n,k}$ representing the CEEs, the channel with CEEs can be written as
\begin{align} \label{eq:3}
{g}_{l,n,k} = \widehat{g}_{l,n,k}+\widetilde{g}_{l,n,k}.
\end{align}
}
Let $\pl_{i}(t)=\pl_{i,l,n}(t)\pl_{l,n,k}$ denote the overall pathloss of the cascaded links, which can be expressed as
{
\begin{align} \label{pl}
    \pl_{i}(t) = G_{\AP}  G_{{l,n}}(-\vec{r}_{i}) G_{{l,n}}(\vec{r}_{\AP}) G_{i}
    \frac{ A^2 e^{-\kappa_{abs}(f)(d_{i,l}(t)+d_{l})}}{(4 \pi d_{i,l}(t) d_{l})^2}.
\end{align}}
The channel matrix from $\UIRS_l$ to the AP can be denoted as $\vec{G}_l = [\vec{g}_{l,1}, \ldots, \vec{g}_{l,k}, \ldots, \vec{g}_{l,K}]^{\sf H} \in \mathbb{C}^{K \times N}$. The PS matrix of $\UIRS_l$, $\vec{\Theta}_l \in \mathbb{C}^{N \times N}$, is the combination of amplitude and PS of each element of the $l$-th IRS. The PS matrix of the IRS can be described as~\cite{Bjornson_WCL_2020}
\begin{align}\label{eq:5}
    \vec{\Theta}_l = \mathrm{diag}([\kappa_{l,1} e^{j \theta_{l,1}}, ..., \kappa_{l,n} e^{j \theta_{l,n}}, ..., \kappa_{l,N} e^{j \theta_{l,N}}]),
\end{align}
where $\kappa_{l,n} \in [0,1]$ and $\theta_{l,n} \in [0, 2\pi)$ represent the amplitude and the PS of the $\UIRS_{l,n}$, respectively. The cascaded effective channel vector $\vec{h}_{i}^{\UIRS_l} \in \mathbb{C}^{K \times 1}$, from $\uu_i$-to-AP through $\UIRS_l$ can be written as
{
\begin{align}\label{6}
\vec{h}_{i}^{\UIRS_l}(t) 
    &=   (\widehat{\vec{G}}_{l} + \widetilde{\vec{G}}_{l}) 
    \vec{\Theta}_l 
    (\widehat{\vec{g}}_{i,l}(t) + \widetilde{\vec{g}}_{i,l}(t)) =   \underbrace{\widehat{\vec{G}}_{l} \vec{\Theta}_l \widehat{\vec{g}}_{i,l}(t)}\limits_{\widehat{\vec{h}}_{i}^{\UIRS_l}(t)}
    +   \underbrace{\widehat{\vec{G}}_{l} \vec{\Theta}_l \widetilde{\vec{g}}_{i,l}(t)
    +   \widetilde{\vec{G}}_{l} \vec{\Theta}_l {\widehat{\vec{g}}}_{i,l}(t)}\limits_{\widetilde{\vec{h}}_{i}^{\UIRS_l}(t)},
\end{align}
where $\widetilde{\vec{h}}_{i}^{\UIRS_l}(t) \sim \mathcal{CN}(\mathbf{0}, \sigma_{{\widetilde{\vec{g}}}_{i,l}}^2 \widehat{\vec{G}}_{l} \widehat{\vec{G}}_{l}^{\sf H} + \sigma_{\widetilde{\vec{G}}_{l}}^2 \Vert {\widehat{\vec{g}}}_{i,l} \Vert^2 \vec{I}_K ) \in \mathbb{C}^{K}$.
}
The channel matrix from the uplink devices to the AP through $\UIRS_l$ can be denoted as $\vec{H}^{\UIRS_l}(t) = [\vec{h}_{1}^{\UIRS_l}(t), \ldots, \vec{h}_{i}^{\UIRS_l}(t), \ldots, \vec{h}_{I}^{\UIRS_l}(t)] \in \mathbb{C}^{K \times I}$. Considering that $x_i$ is a transmitted signal by $\uu_i$, with a transmit power of $p_{i}$.
Assuming there are $L$ IRSs, where we use a single IRS $l$ at a time.
The received signals at the AP through $\UIRS_l$ can be expressed as
{
\begin{align} \label{eq_rx_signal}
\vec{y}_{\AP}^{\UIRS_l} (t)
= \textstyle\sum_{i=1}^I \sqrt{p_{i}} \widehat{\vec{G}}_{l} \vec{\Theta}_l \widehat{\vec{g}}_{i,l}(t) x_{i}(t) + \textstyle\sum_{i=1}^I \sqrt{p_i} \widehat{\vec{G}}_{l} \vec{\Theta}_l \widetilde{\vec{g}}_{i,l}(t)x_{i}(t) + \textstyle\sum_{i=1}^I \sqrt{p_i}\widetilde{\vec{G}}_{l} \vec{\Theta}_l {\widehat{\vec{g}}}_{i,l}(t) x_{i}(t) + \vec{n}(t),
\end{align}
}
where $\vec{n} \in \mathbb{R}^{K \times 1}$ denotes the AWGN vector at AP, and $[\vec{\Theta}_l]_n = \kappa_{l,n} e^{ j\theta_{l,n} }$ is the $n$-th element on the diagonal that runs from the top left to the bottom right of the matrix $\vec{\Theta}_l$. Moreover, the received signal vector at the AP can be written in matrix form as
\begin{align} \label{eq_rx_signal_IRS_matrix}
\vec{y}_{\AP}^{\UIRS_l}(t) = \vec{H}^{\UIRS_l}(t) \sqrt{\vec{P}_{\uu}}\vec{x}_\uu(t) + \vec{n}(t), 
\end{align}
where $\vec{P}_{\uu}\in \mathbb{C}^{I \times I}$ represents the power matrix of uplink devices.
\subsubsection{{Linear detection of the signal at AP}}
Let $\hat{\vec{x}}_{\uu}^{\UIRS_l}(t) \in \mathbb{C}^{I \times 1}$ be the estimated version of $\vec{x}_{\uu}(t)$ using a linear receiver. Considering a linear detection matrix $\vec{V}\in \mathbb{C}^{K \times I}$, which is used to separate the received signal at the AP into $I$ streams as 
\begin{align} \label{eq_rx_d_IRS_matrix}
		\hat{\vec{x}}_{\AP}^{\UIRS_l}(t) &= \vec{V}^{\sf H} \vec{y}_{\AP}^{\UIRS_l} (t) = \vec{V}^{\sf H}\vec{H}^{\UIRS_l} (t) \sqrt{\vec{P}_{\uu}}\vec{x}_\uu (t)  +  \vec{V}^{\sf H} \vec{n} (t).
\end{align}
In multi-user scenarios, optimizing channel gains is no longer the sole objective, as interference must also be considered. While maximizing channel gains is essential for achieving high signal-to-noise ratios, it is equally important to minimize interference levels to ensure efficient and reliable communication among multiple users. This can be achieved through various interference management techniques, such as beamforming, power control, and resource allocation.
Meanwhile, the $i$-th stream of $\hat{\vec{x}}_{\AP}^{\UIRS_l}(t)$, which is used to detect $x_i$, is given by 
{
\begin{align} \label{eq_rx_d_IRS_i}
    \hat{x}_{\AP,i}^{\UIRS_l}(t)  
    &= { \sqrt{p_i}\vec{v}_i^{\sf H}\widehat{\vec{h}}_{i}^{\UIRS_l}(t) x_{i}(t)} +\sqrt{p_i}\vec{v}_i^{\sf H}\widetilde{\vec{h}}_{i}^{\UIRS_l}(t) x_{i}(t)+ {\textstyle\sum_{{i'=1, i'\neq i}}^I \sqrt{p_{i'}}\vec{v}_i^{\sf H}{\vec{h}}_{i'}^{\UIRS_l}x_i'(t)} +{\vec{v}_i^{\sf H}\vec{n}(t)}, 
\end{align}
The interference from other devices and noise is treated as effective noise, and hence, the received signal-to-interference-and-noise ratio (SINR) of $\uu_i$ through $\UIRS_l$ can be expressed as~\cite{liu2020intelligent}
\begin{align} \label{sinr_IRS}
&\gamma_{i}^{\UIRS_l}(t)
   =	\frac{	p_i\|\vec{v}_i^{\sf H} \widehat{\vec{h}}_{i}^{\UIRS_l}(t)|^2 }{ \textstyle\sum_{{i'=1, i'\neq i}}^I  p_{i'}|\vec{v}_i^{\sf H} \widehat{\vec{h}}_{i'}^{\UIRS_l}(t)|^2 + \sum_{i'=1}^{I} p_{i'} \mathbb{E}\{| \vec{v}_{i}^{\sf H} \widetilde{\vec{h}}_{i'}^{\UIRS_l}(t) |^2 \} +\sigma_i^2\Vert\vec{v}_i\Vert^2 }\nonumber \\
   &=   \frac{	p_i\left|\vec{v}_i^{\sf H} \widehat{\vec{h}}_{i}^{\UIRS_l}(t)\right|^2 }{ \textstyle\sum_{{i'=1, i'\neq i}}^I  p_{i'}\left|\vec{v}_i^{\sf H} \widehat{\vec{h}}_{i'}^{\UIRS_l}(t)\right|^2 
   + \sum_{i'=1}^{I} p_{i'} \big[
        \sigma_{\widetilde{\vec{g}}_{i', l}}^2
        \big\Vert \widehat{\vec{G}}_l \vec{v}_{i} \big\Vert^2
        +   \big( \sigma_{\widetilde{\vec{g}}_{i', l}}^2
            \sigma_{\widetilde{\vec{G}}_l}^2 
            + \sigma_{{\widetilde{\vec{G}}_l}}^2 \Vert \widehat{\vec{g}}_{i', l} \Vert^2
        \big)  \Vert \vec{v}_{i} \Vert^2
    \big]
   + \sigma_i^2\Vert\vec{v}_i\Vert^2 }
\end{align} }
{
The achievable uplink rate of $\uu_i$ through $\UIRS_l$ is given by
\begin{align} \label{R_uplink}
	R_{i}^{\UIRS_l}(t)&= \log_2(1+{\gamma}_{i}^{\UIRS_l}.
\end{align}}
The sum rate for the uplink devices through $\UIRS_l$ is the sum of the individual devices' rate through $\UIRS_l$, which can be expressed as 
\begin{align}
    R_{\tt sum}^{\UIRS_l}(t)= \textstyle\sum_{i=1}^I R_{i}^{\UIRS_l}(t).
\end{align}
\subsection{{Downlink Modeling of the Considered IRS-THz Network}}
{Consider the downlink channel from the AP with $K$ antennas to a single antenna downlink device through the $\DIRS$. Let $[\widehat{\vec{g}}_{k,m}]_n =  \widehat{g}_{k,m,n}$ be an individual estimated channel from antenna $k$ of the AP-to-$\DIRS_{m,n}$, where $\widehat{\vec{g}}_{k,m} \in \mathbb{C}^{N \times 1}$ is the estimated channel vector from antenna $k$-to-$\DIRS_m$, which can be written as }
{ $\widehat{g}_{k,m,n} = \sqrt{\pl_{k,m,n}} e^{ - j \omega d_{m,n}},$ where $\pl_{k,m,n}$  represents the pathloss over the link from antenna $k$-to-$\DIRS_{m,n}$. Furthermore, the channel with CEEs $\widetilde{g}_{k,m,n}$ can be expressed as
\begin{align}\label{eq:4.1}
    {g}_{k,m,n} = \widehat{g}_{k,m,n}+\widetilde{g}_{k,m,n}.
\end{align}
The channel matrix from the AP-to-$\DIRS_m$ can be denoted as $\vec{G}_m = [\vec{g}_{m,1}, \ldots, \vec{g}_{m,k}, \ldots, \vec{g}_{m,K}] \in \mathbb{C}^{N \times K}$. The estimated channel from $\DIRS_m$-to-$\du_j$ at time slot $t$ can be represented as $\widehat{\vec{g}}_{j,m}(t) \in \mathbb{R}^{N \times 1}$. Thus, $[\widehat{\vec{g}}_{j,m}]_n = \widehat{g}_{j,m,n}$ is the individual channel from $\DIRS_{m,n}$-to-$\du_j$, which can be expressed as}
{
$\widehat{g}_{j,m,n}(t) = \sqrt{\pl_{j,m,n}} e^{ - j \omega d_{j,m}},$ where $\pl_{j,m,n}$  represents the pathloss of the $\UIRS_{m,n}$-to-$\du_j$ link. Furthermore, the channel with CEEs can be written as
\begin{align}\label{eq:C5.1}
{g}_{j,m,n}(t) = \widehat{g}_{j,m,n}(t)+\widetilde{g}_{j,m,n}(t)
\end{align}
where $\widetilde{g}_{j,m,n}(t)$ is an estimation error and is uncorrelated with the estimated channel ${g}_{j,m,n}(t)$ and the entries of $\widetilde{\vec{g}}$ are IID complex Gaussian with zero mean and variance $\sigma_{\widetilde{\vec{g}}}^2$. The correlation between the estimated and actual channel CSI, which is assumed to be the same for all gains, is given in~\eqref{cor1}. Let $\pl_{j}=\pl_{j,m,n}\pl_{k,m,n}$ denote the overall pathloss of the cascaded downlink, which is given in~\eqref{pl}.
The cascaded channel vector $\vec{h}_{j}^{\DIRS_m} \in \mathbb{R}^{K \times 1}$ from the AP to $\du_j$ through $\DIRS_m$ can be expressed as~\cite{al2020intelligent}
\begin{align}
\vec{h}_{j}^{\DIRS_m}(t) = 
    \underbrace{\widehat{\vec{G}}_{m}^{\sf H}\vec{\Theta}_m \widehat{\vec{g}}_{j,m}(t)}\limits_{\widehat{\vec{h}}_{j}^{\DIRS_m}(t)}
    +   \underbrace{\widehat{\vec{G}}^{\sf H}_m \boldsymbol{\Theta}_m \widetilde{\vec{g}}_{j, m}
    +   \widetilde{\vec{G}}^{\sf H}_m \boldsymbol{\Theta}_m (\widehat{\vec{g}}_{j, m}+\widetilde{\vec{g}}_{j, m})}\limits_{\widetilde{\vec{h}}_{j}^{\DIRS_m}(t)},
\end{align}
where $\widetilde{\vec{h}}_{j}^{\DIRS_m}(t) \sim \mathcal{CN}(\mathbf{0}, \sigma_{{\widetilde{\vec{g}}}_{j, m}}^2 \widehat{\vec{G}}_{m} \widehat{\vec{G}}_{m}^{\sf H} + \sigma_{\widetilde{\vec{G}}_{m}}^2 \Vert {\widehat{\vec{g}}}_{j, m} \Vert^2 \vec{I}_K ) \in \mathbb{C}^{K}$.
} 
The channel matrix can be written as $\vec{H}^{\DIRS_m}(t) = [\vec{h}_{1}^{\DIRS_m}(t),..,\vec{h}_{j}^{\DIRS_m}(t), \ldots, \vec{h}_{J}^{\DIRS_m}(t)]^{\sf H} \in \mathbb{C}^{J \times K}$.
We consider a precoding matrix $\vec{W} \in \mathbb{C}^{K \times J}$, in which $\vec{W} = [\vec{w}_1,\ldots, \vec{w}_j,\ldots, \vec{w}_J]$, where $\vec{w}_j \in \mathbb{C}^{K \times 1}$ is the precoding vector for $\du_j$, which can be expressed as $\vec{w}_j = \sqrt{p_j}\overline{\vec{w}}_j,$ where $p_j$ and $\overline{\vec{w}}_j$ are the transmit power scaling factor and the beamforming vector for receiver $j$, respectively. For the given power budget $\vec{P}_{\AP}$, the power constraint can be written as
 $ \textstyle \sum_{j=1}^J \Vert\vec{w}_j\Vert^2 = \mathsf{Tr} \{ \vec{W} \vec{W}^{\sf H}\} \leq \vec{P}_{\AP}.$ 
Let $\vec{x}(t) \in \mathbb{C}^{K \times 1}$ be the transmit signal vector from the AP to $\du_J$, where
$x_j(t)$ is the data intended for $\du_j$, then, the transmitted signal vector from the AP is written as 
	 $\vec{x}(t)=\textstyle\sum_{{j=1 }}^J{\vec{w}_j} x_j(t)=\textstyle\sum_{{j=1  }}^J\sqrt{p_j} \overline{\vec{w}}_j x_j(t)$~\cite{hu2020sum}.
\subsubsection{{Linear Precoder at AP}}

The received signals at $\du_j$ through $\DIRS_m$ at time slot $t$ can be expressed as 
{
\begin{align} \label{eq_rx_signal_j}
		 {y}_{j}^{\DIRS_m}(t) 
   ={{\sqrt{p_j}(\widehat{\vec{h}}_{j}^{\DIRS_m}(t))^{\sf H} \overline{\vec{w}}_{j}x_{{j}}(t)}}+{{\sqrt{p_j}(\widetilde{\vec{h}}_{j}^{\DIRS_m}(t))^{\sf H} \overline{\vec{w}}_{j}x_{{j}}(t)}}+\!\!\!{\textstyle \sum\limits_{\substack{j'=1, j'\neq j}}^J\!\!\!{({\vec{h}}_{j}^{\DIRS_m}(t))^{\sf H} \overline{\vec{w}}_{j'}x_{j'}(t)}\sqrt{p_{j'}}}+{n_j(t)},
\end{align}
where $n_j$ is the AWGN with power $\sigma^2_j$ at $\du_j$. For a given beamforming matrix\footnote{{In our model, we consider an IIoT network with slow mobility, meaning IIoT devices are characterized by limited movement and relatively stable channel conditions. We therefore use the whole transport block precoding matrix, which is a suitable assumption.}} $\overline{\vec{W}}$, the received SINR at $\du_j$ through $\DIRS_m$ can be expressed as 
\begin{align} \label{sinr_downlink}
&\gamma_{j}^{\DIRS_m}(t) 
= \frac{p_j|{(\widehat{\vec{h}}_{j}^{\DIRS_m}(t))^{\sf H} \overline{\vec{w}}_{j}}|^2}{\sum_{{j'=1, j'\neq j}}^J p_{j'} | {(\widehat{\vec{h}}_j^{\DIRS_m}(t))^{\sf H} \overline{\vec{w}}_{j'}}|^2 
+ \sum_{{j'=1}}^J p_{j'} \mathbb{E}\{ | {(\widetilde{\vec{h}}_j^{\DIRS_m}(t))^{\sf H} \overline{\vec{w}}_{j'}}|^2 \} + \sigma^2_j}, \nonumber\\
&= \frac{p_j\left|{(\widehat{\vec{h}}_{j}^{\DIRS_m}(t))^{\sf H} \overline{\vec{w}}_{j}}\right|^2}{\sum\limits_{j' \ne j}^J p_{j'} \left| {(\widehat{\vec{h}}_j^{\DIRS_m}(t))^{\sf H} \overline{\vec{w}}_{j'}}\right|^2 
+ \sum\limits_{j'=1}^{J} p_{j'} \big[
    \sigma_{\widetilde{\vec{g}}_{j, m}}^2
    \big\Vert \widehat{\vec{G}}_m \overline{\vec{w}}_{j'} \big\Vert^2
+   \big( \sigma_{\widetilde{\vec{g}}_{j, m}}^2
        \sigma_{\widetilde{\vec{G}}_{m}}^2 
        + \sigma_{{\widetilde{\vec{G}}_m}}^2 \Vert \widehat{\vec{g}}_{j, m} \Vert^2
    \big)  \Vert \overline{\vec{w}}_{j'} \Vert^2
\big] + \sigma^2_j}
\end{align}
}
{The achievable downlink rate $R_{j}^{\DIRS_m}$ at $\du_j$ through $\DIRS_m$ can be expressed as
\begin{align}
    &R_{j}^{\DIRS_m}(t)=\log_2(1+{\gamma}_{j}^{\DIRS_m}(t)).
\end{align}}
The sum rate of the downlink devices through $\UIRS_m$ is the sum of the individual devices' rates, which can be expressed as
\begin{align}
   R_{\tt sum}^{\DIRS_m}(t)= \textstyle\sum_{j=1}^J R_{j}^{\DIRS_m}(t).
\end{align}
\subsection{Mobility Modeling for The Considered Network}
Mobility models are developed to characterize the movement and mobility of devices as they travel, particularly the changes in their location. These models are classified into two categories: individual movement models and group movement models~\cite{shakhatreh2022mobile}. We consider the random waypoint mobility (RWM) for moving $\uu_h$ and $\du_h$ as a group head, in which devices move randomly within a specified area. IIoT networks involving freely moving nodes, such as those on a factory floor or in a warehouse, can be simulated using this mobility model. Moreover, the remaining $\uu$s and $\du$s (group members) follow the respective group head based on the reference point group mobility (RPGM)~\cite{Wang_TCCN2020}. As a result of this model, nodes travel to a reference point at a fixed speed, stay there for a fixed amount of time, and then move to the next reference point. Using this mobility model, one can simulate IIoT networks in which nodes move in coordinated ways, such as in a manufacturing plant with robots. In the RWM model, the velocity $\nu_h(t) \in [\nu_{min},\nu_{max}]$, where $h\in\{\uu_h, \du_h\}$ and the direction $\phi_h(t)\in [0,2\pi]$ of the group head at time $t$ are chosen based on a memoryless random process. In addition, for the given interval of data collection $T$, we have
\begin{align}
    \sin{(\phi_h(t))}=\frac{y_h(t+T)-y_h(t)}{T \nu_h(t)},\quad\quad
    \cos{(\phi_h(t))} = \frac{x_h(t+T)-x_h(t)}{T \nu_h(t)}.
\end{align}
 \begin{figure}[!htp] 
 \centering \includegraphics[width=0.4\linewidth]{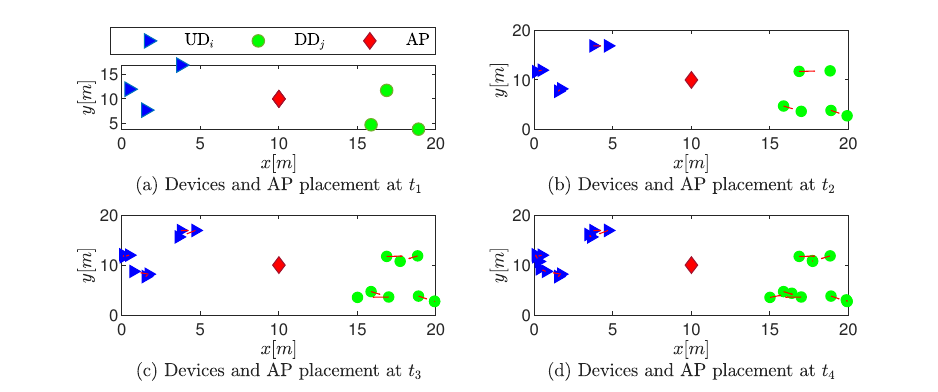} \caption{Device and the AP placement at $t_1$ to $t_4$ during devices movement.}\label{mobilities} 
 \end{figure}
The angle deviation factor and the speed deviation factor are denoted as $\alpha_a$ and $\alpha_s$, respectively, where $\alpha_a \in (-1,1)$ and $\alpha_s \in (-1, 1)$. Furthermore, the velocity and the average direction of group members can be expressed as $\nu_{{\uu}s/{\du}s}(t)=|\nu_h(t)|+\alpha_{s} \nu_{max}$ and $\phi_{{\uu}s/{\du}s}(t)=\phi_h(t)+\alpha_{a} \nu_{max}.
$ By incorporating these factors, our model can capture more realistic mobility patterns of the group members, which is critical for evaluating the performance of our proposed algorithm. The mobility of devices at $t_1$ to $t_4$ is shown in Fig.~\ref{mobilities}.
\section{Problem Formulation}\label{pform}
In IRS-assisted THz networks, there are several unique characteristics and challenges in terms of network and service management, including limited coverage, interference, and resource allocation. Furthermore, the network topology is changing frequently due to the movement of IIoT devices. Meanwhile, the above-mentioned challenges lead to the need for a robust resource allocation algorithm including power allocation at the AP and IRS association algorithm.
{The achievable rate of the end-to-end transmission $\UIRS_l \to \AP \to \DIRS_m$ can be expressed as 
\begin{align} \label{rate}
		& R_{l,m}(t)=\big(1-\frac{\tau}{T}\big) \min \{R_{\tt sum}^{\UIRS_l}(t),  R_{\tt sum}^{\DIRS_m}(t)\},
\end{align}
where $T$ denotes the coherence interval and $\tau$ denotes the time slots occupied for JIIA.} It is crucial to reduce the association time and employ proper JIIA techniques to maximize the sum rate. By reducing the association time, one can allocate more time for data transmission, which can ultimately improve the overall network sum rate. The reflection optimization at the IRS and the power allocation optimization at the AP are used to reduce interference and power consumption, respectively. Additionally, to maximize the sum rate, proper JIIA techniques must be employed to improve channel gain and extend the coverage. 
Let $\vec{\Psi}_{L\times M} = [\Psi_{l,m}]$ be the association matrix, with  $\Psi_{l,m} = 1$ if $\UIRS_l$ is paired with $\DIRS_m$ and $\Psi_{l,m} = 0$ otherwise. In addition, let $\vec{P}_{1\times J}=[p_j]$ be the transmit power vector of the downlink devices. Then, we can establish the optimization problem as follows.
\begin{subequations}\label{eq_opt_prob}
\begin{alignat}{2}
&  (\vec{P}) \text{ : } \underset{\vec{\Theta_\varsigma}\vec{V},\vec{P},\overline{\vec{W}},\vec{\Psi} }{\text{maximize}}
&\quad
& \textstyle \sum_{l,m} R_{l,m} \Psi_{l,m}, \label{eq_optProb}\\
&\quad 
\text{subject to} 
&& \textstyle  {\theta_{\varsigma,n}} \in [0,2\pi), \forall \varsigma, \forall n, \label{eq_constraint1a}\\
&&& \textstyle \kappa_{\varsigma,n} \leq 1, \forall \varsigma, \forall n, \label{eq_constraint1b}\\
&&&  {\Vert v_{i}\Vert} = 1, \forall{i}, \label{eq_constraint2}\\
&&& {\textstyle\sum_{j=1}^{J}p_j\Vert\overline{w}_{j}\Vert^2} \leq  \vec{P}_{\AP}, \label{eq_constraint3a}\\
&&& 
\textstyle  p_{j}\geq 0  , \forall{j}, \label{eq_constraint3b}\\
&&& \textstyle \sum_{l\in L} \Psi_{l,m}\leq 1 , \forall l \in \{1,...,L\}, \label{eq_constraint4}\\
&&& \textstyle \sum_{m\in M} \Psi_{l,m} \leq 1  , \forall m \in \{1, ..., M\}, \label{eq_constraint5}\\
&&& \textstyle  \Psi_{l,m} \in \{0,1\}, \forall l, \forall m, \label{eq_constraint6}
\end{alignat}
\end{subequations}
where $\varsigma\in \{l,m\}$. The objective \eqref{eq_optProb} is the sum rate of the network with IRS reflection $\vec{\Theta}_\varsigma$, given decoding vector $\vec{v_i}$, beamforming vector $\vec{w_j}$, downlink devices transmit power $p_j$, and pairing $\Psi_{l,m}$. Constraints~\eqref{eq_constraint1a} and~\eqref{eq_constraint1b} restrict the PS and amplitude of the reflection model, respectively. Constraints~\eqref{eq_constraint3a} and~\eqref{eq_constraint3a} ensure that the sum of the allocated power cannot exceed the given power budget. Constraint~\eqref{eq_constraint4} ensures that each uplink IRS will be paired to one downlink IRS; constraint~\eqref{eq_constraint5} guarantees that each downlink IRS is allocated to one uplink IRS. Constraints~\eqref{eq_constraint4} and \eqref{eq_constraint5} ensure a one-to-one $\UIRS \to \DIRS$ matching. Additionally, constraint~\eqref{eq_constraint6} ensures the matching variables $\Psi_{l,m}$ are binary. 
\begin{Theorem}\label{T_NP}
    The optimization problem $(\vec{P})$ in \eqref{eq_opt_prob} is an NP-hard problem. Furthermore, the problem $(\vec{P})$ in \eqref{eq_opt_prob} is NP-hard, even if we only consider the IRS association given other variables.
\end{Theorem}
\begin{IEEEproof}
    The proof is given in Appendix~\ref{T_NP_proof}.
\end{IEEEproof}
Since $\vec{P}$ in~\eqref{eq_opt_prob} is NP-hard, solving $\vec{P}$ directly becomes intractable. In the following section, we will reformulate the problem $\vec{P}$ into sub-problems. Then, we develop an optimal solution, specifically, an exhaustive search-based solution for JIIA. Furthermore, a low-complexity algorithm based on matching theory will be proposed for JIIA. 
\subsection{{Problem Decomposition}}
The formulated problem is an MINLP problem because it includes multiple integer variables, e.g., $[\vec{\Psi}]_{l,m} \in \{0,1\}$, and continuous variables $\theta_{\varsigma,n} \in [0, 2\pi]$, and the objective function is nonlinear~\cite{cui2017optimal}. Additionally, the optimization problem in~\eqref{eq_opt_prob} involves optimizing the IRSs' reflection, decoding vector, power allocation, precoding vector, and a matching problem between two disjoint sets (i.e., $\UIRS$s and $\DIRS$s). Thus, our MINLP problem in~\eqref{eq_opt_prob} is NP-hard as it combines optimization over discrete variables with the challenge of dealing with nonlinear functions. Therefore its solution usually involves an enormous search space with exponential time complexity~\cite{jung2013perspective}.

The goal of~\eqref{eq_optProb} is to maximize the sum rate, while \eqref{eq_constraint1a} restricts the PS of IRS elements, and~\eqref{eq_constraint3a} ensures that the sum of allocated power cannot exceed the given power budget. Furthermore, \eqref{eq_constraint4} and \eqref{eq_constraint5} ensure that any $\DIRS$ can be assigned to only one $\UIRS$. This problem is a variant multiple-knapsack (VMK) problem. The VMK problem is an NP-hard optimization problem, meaning that it is computationally intractable to find the optimal solution for large instances of the problem. However, a variety of heuristics and approximation algorithms have been developed to solve the VMK problem within a certain tolerance of the optimal solution. In addition, the VMK problem can be used as a tool for optimization problem decomposition. By decomposing the optimization problem, it can be more tractable to find a solution, as the sub-problems can be solved independently and in parallel. This can also reduce the complexity of the problem, as the sub-problems may be simpler and easier to solve than the original problem. Hence, we can decompose this problem into five sub-problems. Firstly, we rewrite the formulated problem to optimize the PS given the remaining variables as 
    \begin{alignat}{2}\label{eq_opt_prob_ps}
& (\vec{P1}) \text{ : } \underset{{\vec{\Theta}_\varsigma}}{\text{maximize}}
\textstyle (\sum_{l,m} R_{l,m} \Psi_{l,m} | \vec{V}, \vec{P}, \overline{\vec{W}}),
& \quad  \text{subject to} \quad\eqref{eq_constraint1a}, \eqref{eq_constraint1b}.      \end{alignat}
Then, the optimization of the decoding variable given other variables can be formulated as
\begin{alignat}{2}\label{eq_opt_prob_v1}
&(\vec{P2})\text{ : } \underset{\vec{V}}{\text{maximize}}
 \textstyle ( \sum_{l,m} R_{l,m} \Psi_{l,m} | \vec{\Theta}_\varsigma^\star, \vec{P}, \overline{\vec{W}} ), 
& \quad  \text{subject to} \quad
 \eqref{eq_constraint2},
\end{alignat}
where $\vec{\Theta}_\varsigma^\star$ is the optimal solution to problem $(\vec{P1})$. In addition, the optimization of downlink devices' power allocation given other variables can be expressed as
\begin{alignat}{2}\label{eq_opt_prob_p}
&(\vec{P3})\text{ : } \underset{\vec{P}}{\text{maximize}}
 \textstyle ( \sum_{l,m} R_{l,m} \Psi_{l,m} | \vec{\Theta}_\varsigma^\star, \vec{V}^\star, \overline{\vec{W}} ), 
& \quad  \text{subject to} \quad
 \eqref{eq_constraint3a}, \eqref{eq_constraint3b},
\end{alignat}
where $\vec{V}^\star$ is the optimal decoding matrix to problem $(\vec{P2})$. Furthermore, we rewrite the optimization problem to optimize the beamforming vector given other variables as
\begin{alignat}{2}\label{eq_opt_prob_w}
&(\vec{P4})\text{ : } \underset{\overline{\vec{W}}}{\text{maximize}}
\textstyle ( \sum_{l,m} R_{l,m} \Psi_{l,m} | \vec{\Theta}_\varsigma^\star, \vec{V}^\star, \vec{P}^\star ), 
& \quad  \text{subject to} \quad
\eqref{eq_constraint3a}, \eqref{eq_constraint3b},
\end{alignat}
where ${\vec{P}}^\star$ is the optimal solution to problem $(\vec{P3})$. Finally, the association of uplink and downlink IRSs given other variables can be formulated as
\begin{alignat}{2}\label{eq_opt_prob_asso}
&(\vec{P5})\text{ : } \underset{{\vec{\Psi}}}{\text{maximize}}
\textstyle ( \sum_{{l,m}} R_{l,m} \Psi_{l,m} | \vec{\Theta}_\varsigma^\star,\vec{V}^\star,\vec{P}^\star,\overline{\vec{W}}^\star ),
& \quad  \text{subject to} \quad
\eqref{eq_constraint4}, \eqref{eq_constraint5}, \eqref{eq_constraint6},
\end{alignat}
where ${\overline{\vec{W}}}^\star$ is the optimal solution to problem $(\vec{P4})$.

\section{{Optimal Solutions to Network Design}} \label{optimal_solution}
{In this section, we find the optimal solutions for the sub-problems to achieve a sub-optimal solution for the original problem $(\vec{P})$.} 
 \subsection{{Phase Shift Configuration $(\vec{P1})$}} \label{optimalref}
 
 Reflection methods are classified as specular reflection and diffuse reflection depending on IRS size and device location. The elements of the IRS act as perfect mirrors in the specular reflection approach according to Snell's reflection law. Meanwhile, the incident wave reflects with the same beam curvature as the incident beam curvature. However, IRS elements scatter the incident wave in all directions in the diffuse reflection approach rather than reflecting with the same beam curvature. For the case of THz communications, specular reflection is unsuitable due to the absorption and scattering losses. Therefore, diffuse reflection is utilized in the considered IIoT networks. Furthermore, we optimize the reflection model to maximize the achievable data rate with fixed transmit power and decoding matrix/precoding matrix~\eqref{eq_opt_prob_ps}. In order to maximize the received SINR,  we need to maximize the power gain by optimizing the reflection coefficient of IRS elements. By utilizing \eqref{eq:2}, \eqref{eq:3}, and \eqref{eq:5}, the estimated channel in \eqref{6} becomes 
\begin{align}\label{6.1}
|\widehat{h}_{i}^{\UIRS_l}|^2 &= \textstyle(\sum_{n=1}^N\sum_{k=1}^K\sqrt{\pl_{i}} e^{ - j \omega d_{l,n,k}}\kappa_{l,n} e^{j \theta_{l,n}}e^{ - j \omega d_{i,l,n}})^2= (\sum_{n=1}^N\sum_{k=1}^K \kappa_{l,n} \sqrt{\pl_{i}} e^{ - j \omega (d_{l,n,k}+d_{i,l,n})+j \theta_{l,n}})^2.
\end{align}
 To make the positive contribution of all the reflecting signals, we optimize the PS by setting $\theta_{l,n}=\omega (d_{l,n,k}+d_{i,l,n})$ and $\kappa_{l,n}=1$. By doing this, \eqref{6.1} is written as $|\widehat{h}_{i}^{\UIRS_l}|^2 = (\textstyle\sum_{n=1}^N\sum_{k=1}^K \sqrt{\pl_{i}})^2.$
We can achieve the optimal solution for the downlink case by following the same procedure. Furthermore, the optimal PS and amplitude of reflecting elements at the $\varsigma$-th IRS can be expressed as $\theta^\star_{\varsigma,n} = \omega(d_{i,\varsigma}+d_\varsigma),$ and
$\kappa^\star_{\varsigma,n} = 1, \forall{n},$ respectively. Thus, the optimal PS configuration at the $\varsigma$-th IRS is determined as
\begin{align} \label{eq_opt_Theta}
    \vec{\Theta}_\varsigma^\star = \mathrm{diag}[\kappa^\star_{\varsigma,1} e^{ j\theta^\star_{\varsigma,1}},.., \kappa^\star_{\varsigma,n} e^{ j\theta^\star_{\varsigma,n}},..,\kappa^\star_{\varsigma,N} e^{ j\theta^\star_{\varsigma,N}}].
\end{align}
\subsection{{Minimum Mean-Square Error (MMSE) Receiver $(\vec{P2})$}}
Here, we find the optimal decoder with fixed transmit power for all uplink devices and the obtained optimal PS matrix ($\Theta^\star$). The MMSE can be used in a variety of IIoT network applications, including channel estimation, equalization, and detection. In detection, MMSE can be used to minimize the mean square error between estimated and actual transmitted signals in the presence of noise and interference. IIoT networks often involve a large number of sensors and devices communicating over wireless channels in noisy environments, making MMSE particularly useful. Therefore, MMSE can improve the accuracy and reliability of communication systems by reducing the effects of noise and interference.
For the given transmit power, the SINR of all uplink devices is independently maximized by minimizing the respective MMSE between the transmitted and estimated signals~\cite{schubert2005iterative}. As a result, the decoding matrix can be expressed as
\begin{align} \label{mmse_1}
		\vec{V}^{\star} &= \underset{{\vec{V}}}{\mathrm{argmin}}\hspace{1pt} \mathbb{E}\{||
   \vec{\hat{x}}_{\AP}^{\UIRS_l}
   - \vec{x}_{\rm UU}||^2\}=\underset{{\vec{V}}}{\mathrm{argmin}}\hspace{1pt} \mathbb{E}\{||
   \vec{V}^{\sf H}\vec{y}_{\AP}^{\UIRS_l}
   - \vec{x}_{\rm UU}||^2\}  =\underset{{\vec{V}}}{\mathrm{argmin}}\hspace{1pt} \textstyle\sum_{{\hat{i}=1}}^I \mathbb{E}\{|
   \vec{v}_i^{\sf H}\vec{y}_{\AP}^{\UIRS_l}
   - \vec{x}_{i}|^2\}.
\end{align}

Moreover, the decoding vector for $\uu_i$ can be expressed as 
{
\begin{align} \label{mnse_r1} 
   \vec{v}_{i}^\star &= \underset{{\vec{v}_i}}{\mathrm{argmin}} \hspace{1pt} \mathbb{E}\{|\underbrace{
   \vec{v}_i^{\sf H} \vec{y}_{\AP}^{\UIRS_l}
   }_{\hat{x}_i| \vec{v}_i}
   - x_i|^2\}
   =\!\!(\sigma^2_{\widetilde{\vec{g}}_{i,l}} \widehat{\vec{G}}_l \widehat{\vec{G}}_l^{\sf H} + \sigma^2_{\widetilde{\vec{G}}_{l}} \Vert \widehat{\vec{g}}_{i, l} \Vert^2 \vec{I}_K )
    + \sum_{{i'=1, i'\neq i}}^I  p_{i'}\vec{h}_{i',l} \vec{h}_{i',l}^{\sf H} 
\Big)^{
		-1}\widehat{\vec{h}}_{i,l},
\end{align}}
where $\vec{I}_K$ denotes the $K \times K$ identity matrix. Thus, the SINR $(\gamma^\star)$ can be expressed as
{
\begin{align} \label{sinr_mmse_IRS}
& \gamma^{\star}_{\UIRS_l,i} = 
p_i \widehat{\vec{h}}_{i,l}^{\sf H} 
\Big( \sigma_i^2\vec{I}_K
    +   \sum_{i=1}^{I} p_i (\sigma^2_{\widetilde{\vec{g}}_{i,l}} \widehat{\vec{G}}_l \widehat{\vec{G}}_l^{\sf H} + \sigma^2_{\widetilde{\vec{G}}_{l}} \Vert \widehat{\vec{g}}_{i, l} \Vert^2 \vec{I}_K )
    + \sum_{{i'=1, i'\neq i}}^I  p_{i'}\vec{h}_{i',l} \vec{h}_{i',l}^{\sf H} 
\Big)^{
		-1}\widehat{\vec{h}}_{i,l},
\end{align}
}
{\subsection{{Power Allocation Optimization $(\vec{P3})$}}
Power allocation affects IIoT network performance and ultimately impacts the quality of service (QoS) provided to end users. It is essential to optimize various network resources, including power allocation, in order to ensure effective network and service management. Furthermore, power allocation is a critical resource that must be efficiently managed to maximize network performance while maintaining QoS standards.
In this section, we present a power allocation optimization algorithm for the AP that uses the Karush-Kuhn-Tucker (KKT) approach. It is possible to perform this optimization in real-time, which enables the network to adapt to changing traffic and environmental conditions in real-time.
The optimization problem $(\vec{P3})$ to maximize the sum rate through power allocation can be expressed as 
\begin{subequations}\label{eq_opt_P3}
\begin{align}\label{eq_opt_P3_obj}
  (\vec{P3}) \text{ : } &\underset{\vec{P}}{\text{maximize}}
 \sum_{j=1}^{J} \log_2\Big(1+\frac{p_j\left|\widehat{h}_{j,j}(t)\right|^2}{\textstyle\sum_{{j'=1, j'\neq j}}^J p_{j'} \left|h_{j,j'}(t) \right|^2+p_j\mathcal{X} +\sigma^2_j}\Big),\\ &
\text {subject to} \quad
 {\textstyle\sum_{{j=1}}^J p_j\Vert\overline{w}_{j}\Vert^2} \leq  \vec{P}_{\AP}, \quad\quad\textstyle  p_{j}\geq 0  , \quad\quad\forall{j},\label{eq_constraint_P3_1}
\end{align}
\end{subequations}
where $\widehat{h}_{j,j}(t)$  is the effective channel gain of the desired device's signal, ${h}_{j,j'}(t)$ is the composition of effective channel gain and estimation error of the $j'$-th device's signal, which is an interference for the desired device, and $\mathcal{X}= \sigma_{\widetilde{\vec{g}}_{j, m}}^2    \Vert \widehat{\vec{G}}_m \overline{\vec{w}}_{j} \Vert^2
+ ( \sigma_{\widetilde{\vec{g}}_{j, m}}^2\sigma_{\widetilde{\vec{G}}_{m}}^2 +\sigma_{{\widetilde{\vec{G}}_m}}^2 \Vert \widehat{\vec{g}}_{j, m} \Vert^2)  \Vert \overline{\vec{w}}_{j} \Vert^2$. The objective of this section is to offer a closed-form solution to the power allocation problem that supports the offline calculation of power coefficients using the channel's CSI. The following theorem presents the closed-form solution to~\eqref{eq_opt_P3}.} 
\begin{Theorem}\label{T_P}
{The power allocation matrix is written as $\vec{P}_{1\times J}= \mathsf{diag}\{p_1,\ldots,p_j,\ldots,p_J\}$, and $p_j$ is analytically determined by the water-filling solution as
\begin{align} \label{L3_1}  
p_j = 
\left[
    \frac{\frac{1}{\mu \Vert\overline{\vec{w}}_{j}\Vert^2+\upsilon_j}-\frac{ \iota_j(t)+\sigma^2_j}{\left|\widehat{h}_{j,j}(t) \right|^2}}{1+\frac{\sigma_{\widetilde{\vec{g}}_{j, m}}^2    \Vert \widehat{\vec{G}}_m \overline{\vec{w}}_{j} \Vert^2
+ ( \sigma_{\widetilde{\vec{g}}_{j, m}}^2\sigma_{\widetilde{\vec{G}}_{m}}^2 +\sigma_{{\widetilde{\vec{G}}_m}}^2 \Vert \widehat{\vec{g}}_{j, m} \Vert^2)  \Vert \overline{\vec{w}}_{j} \Vert^2}{\left|\widehat{h}_{j,j}(t) \right|^2}}
\right]^{+},
\end{align}
where $[x]^+ = \max \{x,0\}$, $\iota_j(t)$ is the interference term, and $\mu$ and $\upsilon_j$ are the Lagrangian multipliers. The value of $\mu$ is selected to satisfy ${\textstyle\sum_{{j=1}}^J p_j\Vert\overline{w}_{j}\Vert^2} =  \vec{P}_{\AP},$ and $\upsilon$ is selected to ensure a non-negative value.}
\end{Theorem}
\begin{IEEEproof}
The proof is given in Appendix~\ref{T_P_proof}.
\end{IEEEproof}
{We invoke Theorem~\ref{T_P} to obtain the optimal $p_j$ and then iterate the water-filling algorithm over all devices. The iterative water-filling algorithm for~\eqref{eq_opt_prob_p} is shown in Algorithm~\ref{algo_p}.}

\begin{algorithm}[!htp]
{\caption{Iterative Water-Filling Algorithm for $\vec{P3}$ in~\eqref{eq_opt_prob_p} }\label{algo_p}
 {\bf Input: } $J$, power allocation matrix $\vec{P}_{1\times J}=[p_j]$, convergence threshold $\epsilon > 0$\\
{\bf Initialize: } $p_j(0) = \vec{P}_{AP}/J$,  $n=0$ \\
\For {$j \in J$}{
 \While{($p_j(n)-p_j(n-1) > \epsilon$)} {
 $\hat{p}_j \gets p_j(n)$\\
 $n=n+1$ \\
 \While{(until $\hat{p}_j$ is not converged)} {
$\iota_j= \sum\limits_{j'=1, j' \ne j}^J\Big( p_{j'} \left| {(\widehat{\vec{h}}_j^{\DIRS_m}(t))^{\sf H} \overline{\vec{w}}_{j'}}\right|^2 
+  \big[
    \sigma_{\widetilde{\vec{g}}_{j, m}}^2
    \big\Vert \widehat{\vec{G}}_m \overline{\vec{w}}_{j'} \big\Vert^2
+   \big( \sigma_{\widetilde{\vec{g}}_{j, m}}^2
        \sigma_{\widetilde{\vec{G}}_{m}}^2 
        + \sigma_{{\widetilde{\vec{G}}_m}}^2 \Vert \widehat{\vec{g}}_{j, m} \Vert^2
    \big)  \Vert \overline{\vec{w}}_{j'} \Vert^2
\big]\Big) $\\
Obtain $\upsilon_j$ from equation (13) in~\cite{lee2008simplified}\\
Obtain $\mu$ via bisection on $\vec{P}_{\AP}=\sum_{j=1}^{J}\left[{\Vert\overline{w}_{j}\Vert^2}\Bigg(\frac{\frac{1}{\mu \Vert\overline{w}_{j}\Vert^2+\upsilon_j}-\frac{ \iota_j(t)+\sigma^2_j}{\left|\widehat{h}_{j,j}(t) \right|^2}}{1+\frac{\sigma_{\widetilde{\vec{g}}_{j, m}}^2    \Vert \widehat{\vec{G}}_m \overline{\vec{w}}_{j} \Vert^2
+ ( \sigma_{\widetilde{\vec{g}}_{j, m}}^2\sigma_{\widetilde{\vec{G}}_{m}}^2 +\sigma_{{\widetilde{\vec{G}}_m}}^2 \Vert \widehat{\vec{g}}_{j, m} \Vert^2)  \Vert \overline{\vec{w}}_{j} \Vert^2}{\left|\widehat{h}_{j,j}(t)\right|^2}}\Bigg)\right]^{+} $ \\
$p_j=\left[\frac{\frac{1}{\mu \Vert\overline{w}_{j}\Vert^2+\upsilon_j}-\frac{ \iota_j(t)+\sigma^2_j}{\left|\widehat{h}_{j,j}(t) \right|^2}}{1+\frac{\sigma_{\widetilde{\vec{g}}_{j, m}}^2    \Vert \widehat{\vec{G}}_m \overline{\vec{w}}_{j} \Vert^2
+ ( \sigma_{\widetilde{\vec{g}}_{j, m}}^2\sigma_{\widetilde{\vec{G}}_{m}}^2 +\sigma_{{\widetilde{\vec{G}}_m}}^2 \Vert \widehat{\vec{g}}_{j, m} \Vert^2)  \Vert \overline{\vec{w}}_{j} \Vert^2}{\left|\widehat{h}_{j,j}(t)\right|^2}}\right]^{+} $\\
Normalize to satisfy the power budget\\
$p_j(n) \gets \hat{p}_j$
 } 
 }
 }
 {\bf Output: } {optimal power allocation matrix $\vec{P}^\star$} }
\end{algorithm}

\subsection{Minimum Mean-Square Error (MMSE) based beamforming $(\vec{P4})$ for DL transmission}

The MMSE beamforming has similar properties as the MMSE receiver; thus, the SINR at downlink device $j$ is given as
{
\begin{align}
\gamma_{\DIRS_m,j}^{\star} =p_{j} \widehat{\vec{h}}_{j,m}^{\sf H} 
\left( 
    \sigma_j^2\vec{I}_K  
    +   \sum_{j=1}^{J} p_j (\sigma^2_{\widetilde{\vec{g}}_{j, m}} \widehat{\vec{G}}_m \widehat{\vec{G}}_m^{\sf H} + \sigma^2_{\widetilde{\vec{G}}_{m}} \Vert \widehat{\vec{g}}_{j, m} \Vert^2 \vec{I}_K )
    + \sum_{{j'=1, j'\neq j}}^J  p_{j'}\vec{h}_{j',m} \vec{h}_{j',m}^{\sf H}
\right)^{
		-1}  \widehat{\vec{h}}_{j,m}.
\end{align}}
The MMSE beamforming technique is designed to effectively balance the mitigation of interference to other users and the Gaussian background noise. It achieves optimal performance by maximizing the SINR regardless of the SNR value. Depending on the level of interference, the MMSE beamformer exhibits characteristics similar to both the zero-forcing beamformer (ZFBF) when dealing with high inter-user interference, and the maximum ratio transmission (MRT) beamformer when the interference is low.
\subsection{{Optimal JIIA Strategy using Exhaustive Search (ES) $(\vec{P5})$}}\label{es}
In this subsection, we focus on the IRS association problem in~\eqref{eq_opt_prob} to find the optimal association matrix $\vec{\Psi}$. In our ES method, the $\UIRS \to \DIRS$ sum rate is utilized to find the optimal pairing by checking all possible combinations and selecting the optimal one~\cite{wang2020channel,burer2012non}.
Thus, our model has $L!$ $\UIRS$ to $\DIRS$ combinations. The ES method for~\eqref{eq_opt_prob} is shown in Algorithm~\ref{algo_es}. The association matrix and the achievable rate are set to zero at the beginning of the algorithm. Furthermore, we check the sum rate of all possible combinations; if the sum rate of the current combination is greater than the existing combination, we select the current combination; otherwise, we keep the existing combination. This process is iterated until we check all combinations, as shown in steps $4$ to $10$ of Algorithm~\ref{algo_es}, thus reaching the optimal association matrix $\vec{\Psi}^\star$ for JIIA in~\eqref{eq_opt_prob_asso}.
\begin{algorithm}[!tp]
\caption{Exhaustive Search-based JIIA Algorithm for $\vec{P5}$ in~\eqref{eq_opt_prob_asso} }\label{algo_es}
 {\bf Input: } $L$, $M$, $R_{l,m}$, permutation of set $L$ as $\vec{C}_{\tt temp}$, association matrix $[\vec{\Psi}]_{l,m}$\\
{\bf Initialize: } $[\vec{\Psi}]_{l,m} = \emptyset$,  $\sum R_{l,m}=0$ \\
\For {$i \in L!$}{
 pairing $\UIRS_l$ with $l$-th element of the $i$-th row in $\vec{C}_{\tt temp}$ as $\UIRS_l \gets \vec{C}_{\tt temp}(i,k)$,\quad
 $\vec{C}_{\tt temp}(i,k)$  gives the $\DIRS_m$ \\
 \For {$ l \in L$}{
 $\sum R_{l,m}$ as in equation~\eqref{rate}
 }
 {\bf if} {($\sum R_{l,m} \geq \sum R_{l,m}^\star$ 
 )}\\
 {
 $[\vec{\Psi}]_{l,m}^\star  \gets [\vec{\Psi}]_{l,m} $\\ 
 {\bf else}\\
 $[\vec{\Psi}]_{l,m}^\star  \gets [\vec{\Psi}]_{l,m}^\star $\\
 }
 }
 {\bf Output: } {optimal association matrix $[\vec{\Psi}]^\star_{l,m}$} 
\end{algorithm}
$(\vec{P5})$ in~\eqref{eq_opt_prob_asso}.

\section{The Proposed Matching-based JIIA Algorithm}\label{proposed}
IRS-aided networks require a massive number of reflecting elements $N$, especially at higher frequencies, where the number of elements can reach $100 \times 100$. The ES method is not an efficient scheme for association problems. Therefore, to optimize the association problem in~\eqref{eq_opt_prob_asso}, we propose a Gale-Shapley algorithm-based~\cite{gale1962college} solution. {In addition, we assume that the CSI available at the AP is imperfect. During the signaling phase, priority matrix configuration and matching are done by the AP based on the CSI and the JIIA decision is then broadcast to all IIoT devices for the data transmission phase. The matching algorithm is presented in detail in the following subsections.}
\subsection{Rules for Matching}
In this subsection, we introduce some matching rules for the formulated problem in~\eqref{eq_opt_prob_asso}.
\begin{Definition}\textit{\textbf{(Preference Rule)}} \label{def_pref}
The $l$-th uplink IRS $\UIRS_l$ prefers the $m$-th downlink IRS $\DIRS_m$ to the $m^{\star}$-th downlink IRS $\DIRS_{m^\star}$ is denoted as  $\DIRS_m \succ^{{\UIRS_l}} \DIRS_{m^\star}$.
\end{Definition}
\begin{Definition}\textit{\textbf{(Propose Rule)}} \label{def_prop}
The uplink IRS $l\in L$ proposes to its most preferred downlink IRS $\DIRS_m$ in its preference list, i.e., $\DIRS_m \succ^{{\UIRS_l}} \DIRS_{m^\star}$.
\end{Definition}
\begin{Definition}\textit{\textbf{(Reject Rule)}} \label{def_reject}
The downlink IRS $m\in M$ rejects the proposing uplink IRS if a better matching uplink IRS exists. Otherwise, the {proposed} uplink IRS that is not rejected will be retained as a matching candidate.
\end{Definition}
\begin{Definition}\textit{\textbf{(Matching)}} \label{def_match}
For the formulated matching problem $(\mathcal{L},\mathcal{M},[\vec{\Upsilon}]_{l,m}^\UIRS,[\vec{\Upsilon}]_{m,l}^\DIRS)$, where $\mathcal{L}$, $\mathcal{M}$, $[\vec{\Upsilon}]_{l,m}^\UIRS$, and $[\vec{\Upsilon}]_{m,l}^\DIRS$ denote the set of uplink IRSs, the set of downlink IRSs, the priority matrix of uplink IRSs, and the priority matrix of the downlink IRSs, respectively. The matching $\Psi(m)=j$ indicates that the downlink IRS $m$ has been matched to uplink IRS $j$, whereas the matching $\Psi(m)=\emptyset$ implies that the downlink IRS $m$ has not been allocated to any uplink IRS.
\end{Definition}
\subsection{Priority Matrix Configuration}
The formulated problem, i.e., \eqref{eq_opt_prob_asso}, seeks to maximize the sum rate of the uplink and downlink through JIIA. The sum rate between $\UIRS_l$ and $\DIRS_m$ is given in~\eqref{rate}.
The $\UIRS$ priorities are obtained based on the sum rate between each $\DIRS$ and the $\UIRS$. After calculating the sum rate between each $\UIRS$ and all the downlink IRSs, the $\UIRS$ priority matrix is constructed. The priority matrix at $\UIRS$s ($[\vec{\Upsilon}]_{l,m}^\UIRS$)/$\DIRS$s ($[\vec{\Upsilon}]_{m,l}^\DIRS$) is constructed with the highest sum rate offered by the $\DIRS$s/$\UIRS$s at the top and the lowest sum rate offered at the bottom, as shown in steps 8 to 19 of Algorithm~\ref{algo: proposed}. Additionally, the dimension of the $\UIRS$ priority matrix $[\vec{\Upsilon}]_{l,m}^\UIRS$ is $L \times M$ and that of the $\DIRS$ priority matrix $[\vec{\Upsilon}]_{m,l}^\DIRS$ is $M \times L$. The priority relations for the $\UIRS$ to the $\DIRS$ are defined in Definition~\ref{def_pref}.
\subsection{The Proposed JIIA Algorithm}
\begin{algorithm}[!tp]
  \caption{Proposed Matching-based JIIA Algorithm for $\vec{P5}$ in \eqref{eq_opt_prob_asso}}  
  \label{algo: proposed}
  \DontPrintSemicolon{
  \KwIn{ $L$, $M$, $R_{l,m}$, $[\vec{\Upsilon}]_{l,m}^{\UIRS}$, $[\vec{\Upsilon}]_{m,l}^{\DIRS}$, set of unmatched $\UIRS$s $\Pi$, $[\vec{\Psi}]_{l,m}$}
 {\bf Initialize }$[\vec{\Upsilon}]_{l,m}^\UIRS = \emptyset$,
 $[\vec{\Upsilon}]_{l,m}^\DIRS = \emptyset$, $[\vec{\Psi}]_{l,m} = \emptyset$;\;
\underline{\textbf{Priority Matrix Configuration:}}\\
\For {$l\in L$}{
$R_{l,m}, \quad \forall\hspace{2mm} m \in M$ and store in $[\vec{R}]_{l,m}^{\UIRS}$\;
}
$[\vec{R}]_{m,l}^{\DIRS} = [[\vec{R}]_{l,m}^{\UIRS}]^{\sf T}$\;
[$[\vec{R}]_{l,m}^{\UIRS}$,$[\vec{\Upsilon}]_{l,m}^\UIRS$] = sort ($[\vec{R}]_{l,m}^{\UIRS}$,2,descend)\;
[$[\vec{R}]_{m,l}^{\DIRS}$,$[\vec{\Upsilon}]_{m,l}^\DIRS$] = sort ($[\vec{R}]_{m,l}^{\DIRS}$,2,descend)\;
\underline{\textbf{UR - DR Association:}}\\
\While {(either $\Pi$ $\neq \emptyset$ or $\UIRS$s not rejected by all $\DIRS$s)}{
\For {$\UIRS_{l'}\in \Pi$}{
propose to the highest priority $\DIRS$ in $[\vec{\Upsilon}]_{l,m}^{\UIRS}$\;
element of $[\vec{\Psi}]_{l,m} = 1$\;
}
\For {$\DIRS_m\in M$}{
{\bf if} {($\DIRS_m \notin [\vec{\Psi}]_{l,m}$ )}\;
{
$[\vec{\Psi}]_{l,m} \gets [\vec{\Psi}]_{l,m} \cup (\UIRS_{l'},\DIRS_m) $\; 
}
{\bf else if} {($R_{m,l'}^{\DIRS} >  R_{m,l}^{\DIRS}$)}\;{
$[\vec{\Psi}]_{l,m} \gets [\vec{\Psi}]_{l,m} \cup (\UIRS_{l'},\DIRS_m)$ \;
{\bf else}\;
$[\vec{\Psi}]_{l,m} \gets [\vec{\Psi}]_{l,m} \cup (\UIRS_{l},\DIRS_m)$ \;
}
}
  }
{\bf Output: } association matrix $[\vec{\overline{\Psi}}]_{l,m}$
}
\end{algorithm}
We deploy a large number of IRSs in the network to assist the communication and assume that the number of IRSs for uplink communications is the same as for the downlink (i.e., $L=M$) in the considered network. We explore the JIIA algorithm as one-to-one matching in the proposed scheme. Each $\UIRS$ can be allocated to at most one $\DIRS$ in one-to-one matching. Algorithm~\ref{algo: proposed} is illustrated as a series of attempts from $\UIRS$s to $\DIRS$s.
The $\DIRS$ can either be associated with the $\UIRS$ or remain vacant during this process. Each $\UIRS$ proposes pairing with the highest priority $\DIRS$ in their priority list until the $\UIRS$ has either been paired or rejected by all $\DIRS$s. $\UIRS$s rejected by the $\DIRS$ are not allowed to try again for the same $\DIRS$. The $\DIRS$ is immediately engaged if the $\UIRS$ proposes a free $\DIRS$; however, if the $\UIRS$ proposes a $\DIRS$ that is already engaged, the $\DIRS$ compares the new and current $\UIRS$ and gets engaged with the one with the highest data rate link. Thus, if the $\DIRS$  favors the current $\UIRS$, the new proposal will be rejected. Alternatively, if the $\DIRS$ prefers the new proposal, then the $\DIRS$  will break the engagement with the current $\UIRS$ and accept the new $\UIRS$. The process is repeated until all $\UIRS$s are engaged, or all options are considered, as shown in steps 8 to 19 of Algorithm~\ref{algo: proposed}.
\begin{Remark}\label{R2}
The proposed JIIA algorithm terminates in polynomial time, i.e., if there are $L$ $\UIRS$ and $M$ $\DIRS$, then the algorithm is terminated after at most $L \times M$ iterations.
\end{Remark} 
\begin{IEEEproof}
In each iteration, each unmatched $\UIRS$ proposes a $\DIRS$ that it has never explored before. Moreover, the $\DIRS$ accepts or rejects the new proposal according to the status and preference of the $\DIRS$, as shown in steps 8 to 19 of Algorithm~\ref{algo: proposed}. As a result, for $L$ $\UIRS$s and $M$ $\DIRS$s, we have $L \times M$ possible proposals occurring in the proposed algorithm.
\end{IEEEproof}
{\begin{Theorem} \label{T1} 

{The JIIA matrix $\vec{\Psi}^\star$ obtained with the ES method  and $\vec{\overline{\Psi}}$ obtained with the proposed JIIA algorithm are similar and converge to the optimal solution for
$(\vec{P5})$ in~\eqref{eq_opt_prob_asso}.}
\end{Theorem}
\begin{IEEEproof}
The proof is given in Appendix~\ref{T1_proof}.
\end{IEEEproof}}
\subsection{Computational Complexity and Convergence Analysis} \label{complexity}
Let's consider $L=M$ for simplicity and use $L$. Then, the computational complexity of the ES is $\mathcal{O}(L!)$. Meanwhile, the computational complexity of the proposed JIIA algorithm is $\mathcal{O}(L^2)$. To compare the computational complexity of different methods, we take the natural logarithm of the linear computational complexity. The logarithm complexity of the ES is $\mathcal{O}(ln(L!))$, which can be simplified using logarithmic properties and Stirling's formula, $\ln (n!)= n\ln(n)-n$. Accordingly, the logarithm complexity of the ES is $\mathcal{O}(L\ln(L)-L)$. Moreover, the logarithm complexity of the proposed scheme can be expressed as $\mathcal{O}(2\ln{L})$, which is much less than that of the ES scheme.

 \begin{figure}[!htp]  \centering \includegraphics[width=0.6\linewidth]{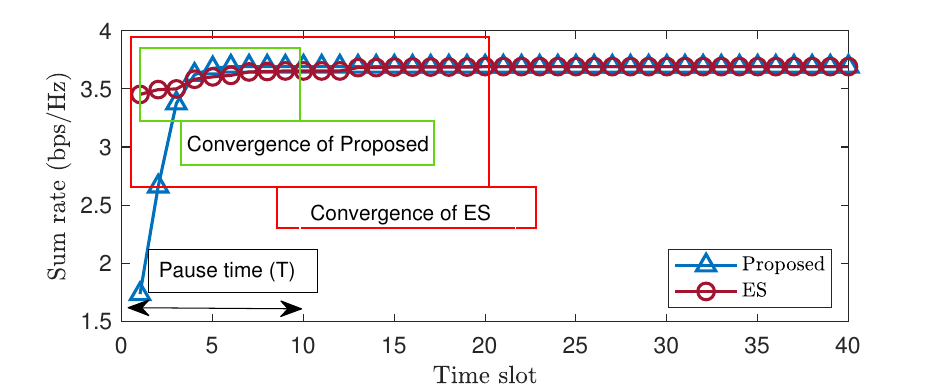} \caption{Sum rate versus  time slot, for a network area of $40\times40$ $\mathrm{m}^2$, $N=100\times100$, and $K=64$.} \label{convergence} \end{figure}
\begin{figure}[!htp] \centering \includegraphics[width=0.6\linewidth]{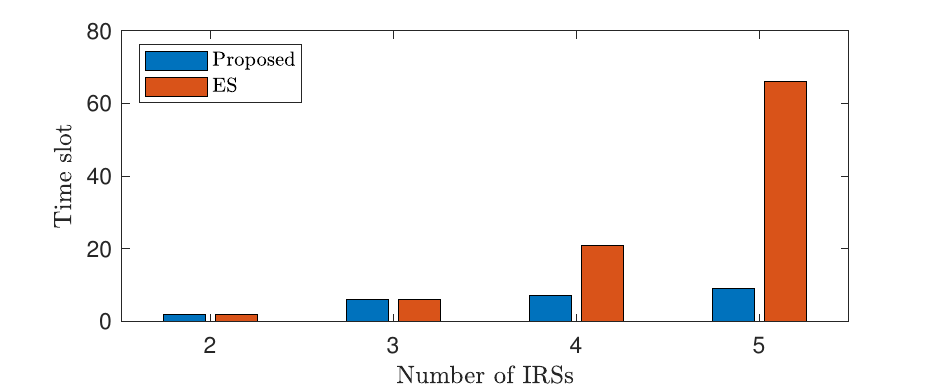} \caption{Convergence rate versus  number of IRSs, for a network area of $40\times40$ $\mathrm{m}^2$, $N=100\times100$, and $K=64$.} \label{CB} \end{figure}
 
Additionally, we consider that the IIoT devices are moving, and the pause time is short. Thus, the ES may not converge during that short time, while our proposed scheme can converge. In addition, it can be observed in Fig.~\ref{convergence} that the proposed scheme converges very fast, while the ES scheme requires more time slots than the pause time to converge. Furthermore, we examine the convergence rate with a given pause time for different numbers of IRSs. It is observed in Fig.~\ref{CB} that both the ES and proposed schemes converge very fast for a small number of IRSs. However, when the number of IRSs increases, the proposed scheme converges in fewer time slots, while the ES requires a huge time to converge, as shown in Fig.~\ref{CB}.

\section{Simulation Results and Discussion} \label{pref}

In our study, MATLAB is used to implement the proposed algorithm. Each point is based on the average of $10^6$ independent trials. Furthermore, we present simulation results evaluating the performance of our proposed algorithm and comparing it to the following schemes as a benchmark. 

\begin{itemize}
    \item \textbf{Random Association (RA)-based JIIA algorithm}: The RA scheme~\cite{cui2017optimal} randomly selects the $\UIRS-\DIRS$ association matrix.
    \item \textbf{Exhaustive Search (ES)-based JIIA algorithm}: In the ES method, the $\UIRS-\DIRS$ sum rate is utilized to find the optimal assignment by checking all possible combinations and selecting the optimal one, as in Section~\ref{es}.
    \item \textbf{Greedy Search (GS)--based JIIA algorithm}: In the GS method~\cite{xu2015user}, each $\UIRS$ selects the $\DIRS$ with the highest data rate. $\DIRS$s receiving multiple proposals are randomly allocated to one of the $\UIRS$s.
    \end{itemize}
\subsection{Simulation Setup}
\begin{figure}[!htp] 
 \centering \includegraphics[width=0.4\linewidth]{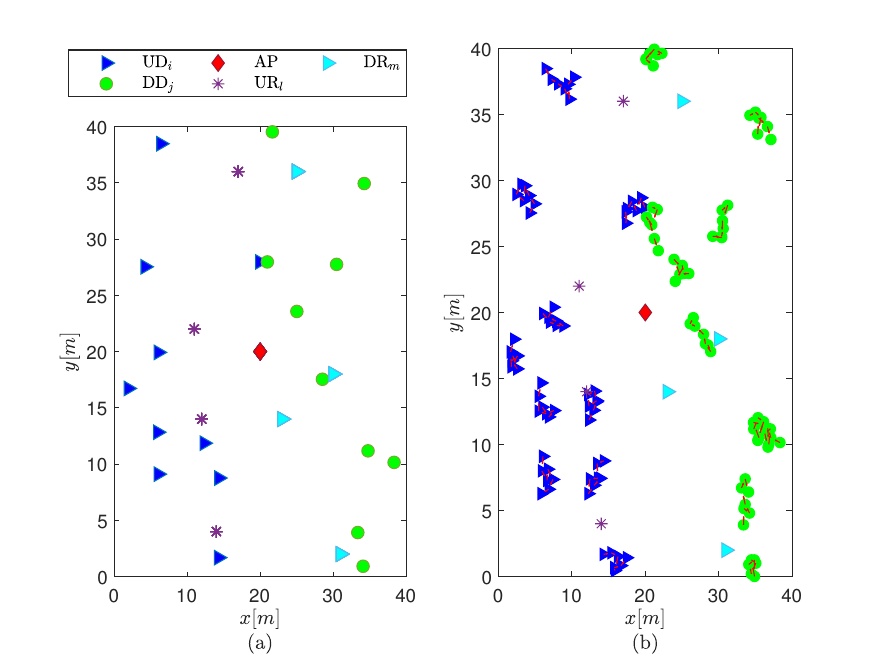} \caption{{Simulation scenario for a network area of $40\times40$ $\mathrm{m}^2$ and $I=J=10$; (a) single snapshot of device deployment, (b) device deployment with mobility.}} \label{T}
 \end{figure}

In the simulated scenario, explained in Section~\ref{model}, {$10$ uplink and downlink devices} are deployed within the network area. Additionally, we deployed IRSs with $100\times 100$ elements at each IRS. The length and width of each IRS element are $N_x=N_y= 0.4 \lambda$~\cite{najafi2020physics}, where $\lambda$ is the wavelength. {Table~\ref{tab: sim} details the simulation parameters based on existing works.} In our study, we used reference point group mobility mode (RPGM) with synchronized velocity and pause time of all devices; however, the directions are different, as shown in Fig~\ref{T}. The snapshot of a single time slot is shown in Fig~\ref{T} (a), i.e., the state of the devices and their positions at a specific moment in time. Furthermore, Fig~\ref{T} (b) depicts the mobility of devices in different directions.
\begin{table}[!htp]
\centering
\renewcommand{\arraystretch}{1}
{\caption{Simulation Parameters}
\label{tab: sim}
\begin{tabular}{l l l l}
\hline
\textbf{Parameters}                                  & \textbf{Values} &\quad\textbf{Parameters}                                  & \textbf{Values} \\ \hline
Carrier frequency, $f_c$ {[}GHz{]}                   & $300$              &\quad
  Number of AP antennas, $K$           & $64$               \\ 
Number of $\du$, $J$                   & $10$              &\quad
     Absorption coefficient, $\kappa_{abs}(f)$ ${[}m^{-1}{]}$   & $0.0033$    
     \\ 
Number of $\uu$, $I$                   & $10$   
&\quad
Noise power density, $N_0$ {[}dBm/Hz{]} & $-174$
\\ 
Transmit power, $P$ {[}dBm{]}                    & $23$ 
& \quad
Network area  [$\mathrm{m}^2$]                            & $40\times40$ 
          \\ 
          Coherence interval                              & $200$  
      & \quad
Number of IRS elements, $N$                              & $100\times100$            \\ 
Channel bandwidth, $B$ {[}GHz{]}                     & $10$
& \quad
Noise figure, NF {[}dB{]}                           & $10$
\\ 
Spacing between IRS elements & $\lambda/2$& \quad
Angle deviation factor, $\alpha_a$ & $(-1,1)$
\\
Spacing between AP antennas & $\lambda/2$
& \quad
Speed deviation factor, $\alpha_s$ & $(-1,1)$
\\
Side length of IRS elements & $0.4 \lambda$
& \quad
$\uu_i$ coordinate $[m]$&  $x_i \sim \mathcal{U}(0\,\,20)$, $y_i \sim \mathcal{U}(0\,\,40)$\\
Amplitude of IRS element coef., $\kappa$ & $1$& \quad
$\du_j$ coordinate $[m]$&  $x_j \sim \mathcal{U}(20\,\,40)$, $y_j \sim \mathcal{U}(0\,\,40)$
\\
PS of IRS elements, $\theta$ & $[0,2\pi]$
& \quad
$\UIRS_l$ coordinate $[m]$&  $x_l \sim \mathcal{U}(5\,\,20)$, $y_l \sim \mathcal{U}(0\,\,40)$\\
Avg. height of IRSs' location $[m]$ & $10$& \quad
$\DIRS_m$ coordinate $[m]$&  $x_m \sim \mathcal{U}(20\,\,35)$, $y_m \sim \mathcal{U}(0\,\,40)$
\\
Avg. height of IIoT devices $[m]$ & $1$
&\quad
$\AP$ coordinate $[m]$&  $[20,20,10]$
\\ \hline
\end{tabular}}
\end{table}
\subsection{Performance Comparison without Association Overhead}
This section evaluates the performance of the proposed algorithm based on the network sum rate without considering the association overhead (i.e., $\tau=0$). The results demonstrate how transmission power, number of antennas at AP, number of IRS elements, and size of the network impact the sum rate. 
\subsubsection{Impact of the transmission power}
\begin{figure}[!htp] 
 \centering \includegraphics[width=0.6\linewidth]{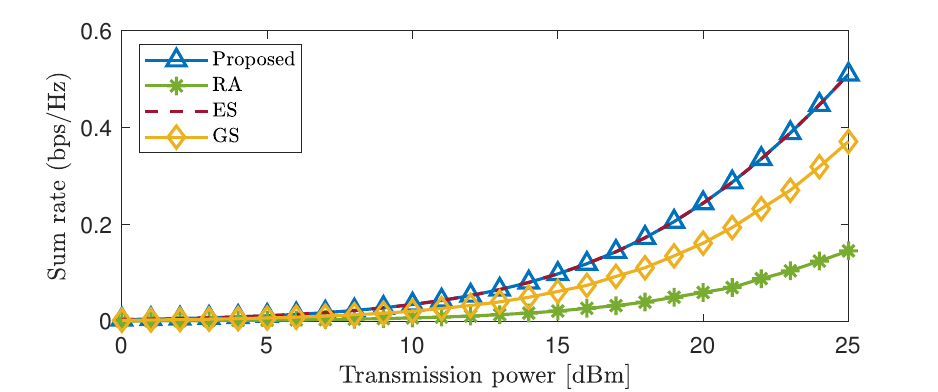} \caption{{Sum rate versus  transmission power, for a network area of $40\times40$ $\mathrm{m}^2$, $N=100\times100$, $I=J=10$, and $K=64$.}} \label{R_Power}
 \end{figure}

{To demonstrate how transmission power affects performance, we use $I=J=10$, $K=64$, $N=100\times 100$, and a network area of $40\times40$ $\mathrm{m}^2$.}  Fig.~\ref{R_Power} shows the sum rate under different transmission powers. For all cases, the sum rate increases with the transmission power. Additionally, the sum rate of the proposed scheme matches that of the ES scheme. Furthermore, the proposed algorithm outperforms the GS and RA algorithms.
\subsubsection{Impact of the number of antennas at the AP}
\begin{figure}[!htp] 
 \centering \includegraphics[width=0.6\linewidth]{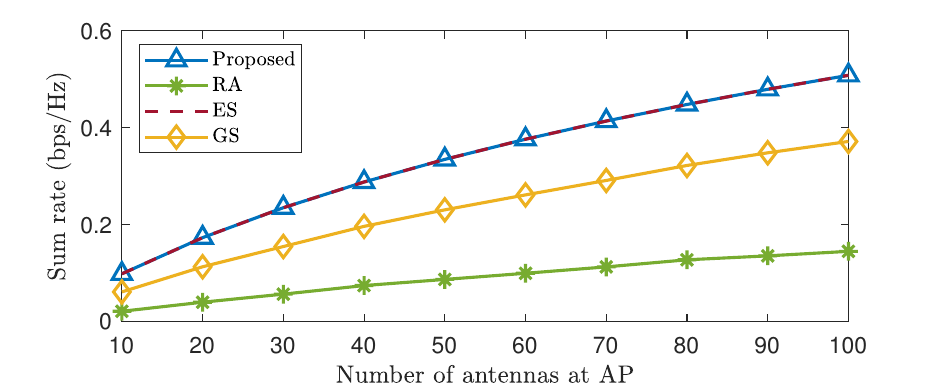}  \caption{{Sum rate versus  number of antennas at the AP, for a network area of $40\times40$ $\mathrm{m}^2$, $N=100\times100$, and $P=23$ dBm.}} \label{R_antenna} 
 \end{figure}
Comparing the performance of conventional large-scale antenna arrays deployed at the AP is imperative. The sum rate increases as the number of antennas at the AP increases, as shown in Fig.~\ref{R_antenna}. This is because when $K$ increases, the total signal gain increases as well, which increases the sum rate. Furthermore, it is observed from Fig.~\ref{R_antenna} that a significant sum rate performance improvement can be achieved by the proposed scheme over the GS and RA schemes.
\subsubsection{Impact of the number of IRS elements}
\begin{figure}[!htp] 
 \centering \includegraphics[width=0.6\linewidth]{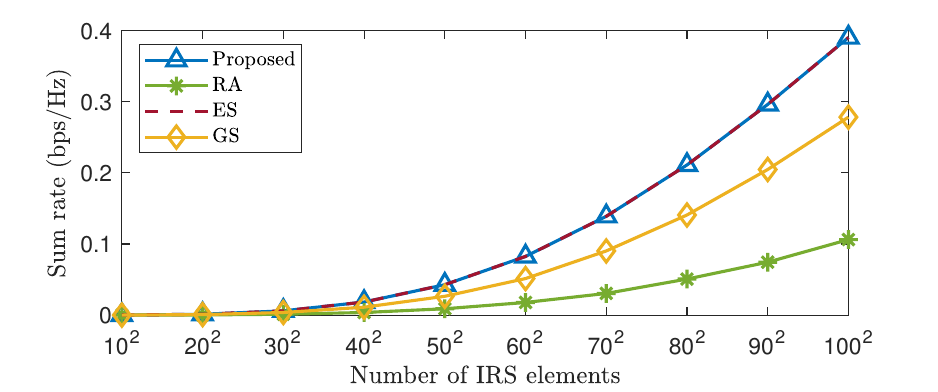}  \caption{{Sum rate versus  number of IRS elements, for a network area of $40\times40$ $\mathrm{m}^2$, $K=64$, $I=J=10$, $K=64$, and $P=23$ dBm.} }\label{R_elements} 
 \end{figure}

In Fig.~\ref{R_elements}, the sum rate is plotted against the number of IRS elements. According to the results, the sum rate of all schemes increases with $N$. This is because the received signal strength increases as $N$ increases, which leads to an improvement in the sum rate. Moreover, the results show that the sum rate of the proposed scheme is similar to that of the ES scheme. Additionally, the proposed scheme achieves a higher sum rate than the GS and RA schemes, due to proper association.
\subsubsection{{Impact of the carrier frequency}}
\begin{figure}[t]
        \centering
        \subfloat[$f_c$= $6$ GHz]{\includegraphics[width=0.33\linewidth]{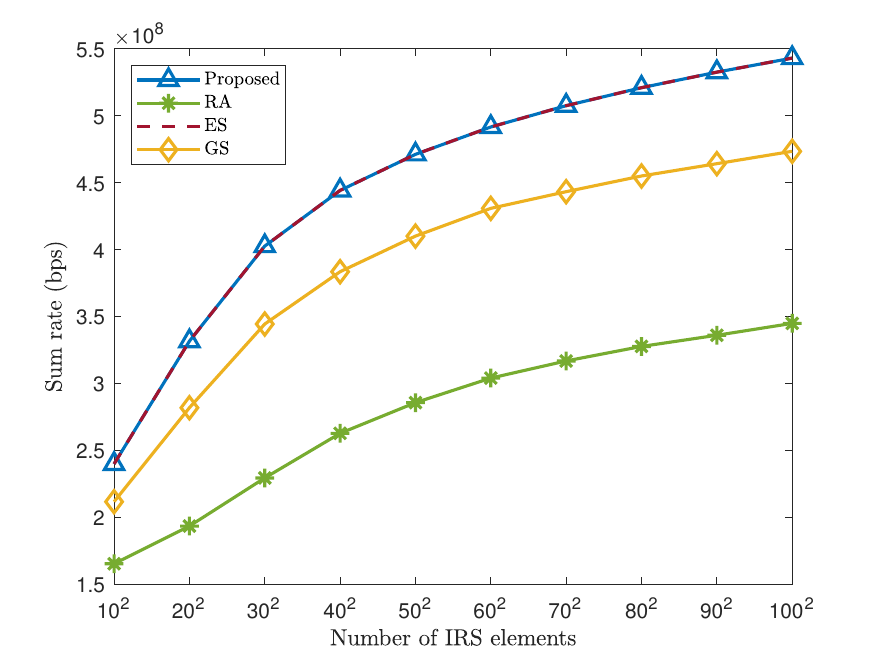}\label{R_6ghz}}
        \subfloat[$f_c$= $28$ GHz]{\includegraphics[width=0.33\linewidth]{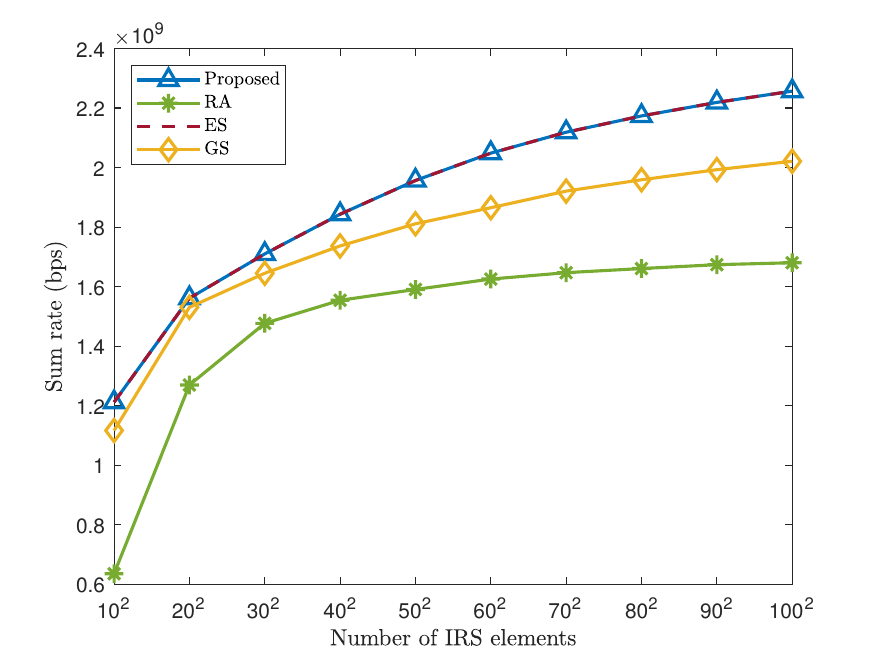}}\label{R_28ghz}
        \subfloat[$f_c$= $300$ GHz] {\includegraphics[width=0.33\linewidth]{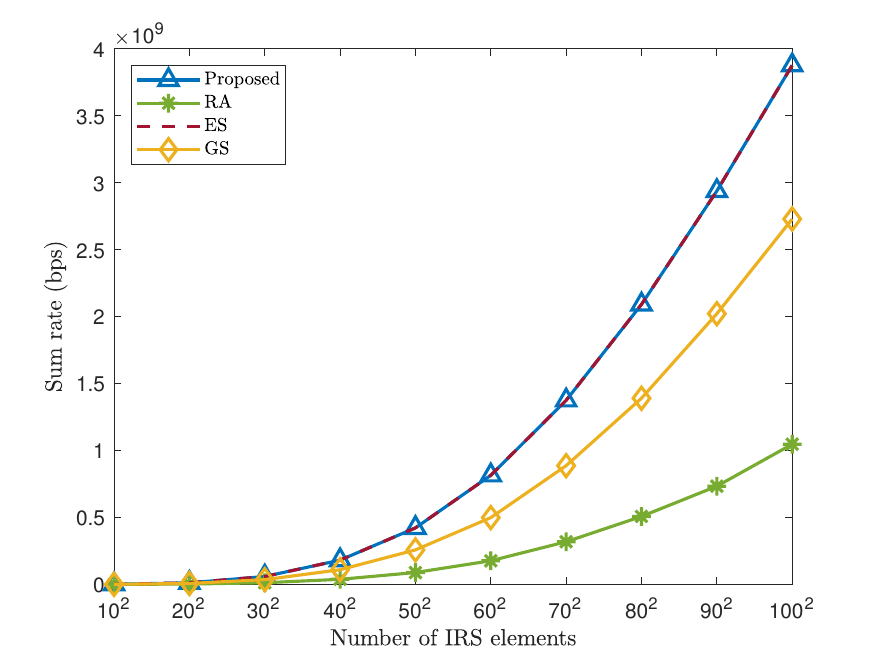}}\label{R_300ghz}
        \caption{{Sum rate versus number of IRS elements, for a network area of $40\times40$ $\mathrm{m}^2$, $K=64$, and $P=23$ dBm.}}
        \label{fig:feq}
\end{figure}

{In Fig.~\ref{fig:feq}, the sum rate is plotted against the number of IRS elements for three different frequencies. The sum rate of all the schemes increases with the number of IRS elements in all three cases. Furthermore, our proposed method achieves a higher sum rate than the GS and RA schemes in all three frequencies. The main purpose of Fig.~\ref{fig:feq} is to compare the sum rates at different frequencies. We therefore measured the sum rate in bits per second (bps) instead of bps/Hz to show the improvement for the THz band. For a fixed number of IRS elements, i.e., $100\times100$, the sum rates achieved at $6$~GHz, $28$~GHz, and $300$~GHz are $0.55$~Gbps, $2.4$~Gbps, and $4$~Gbps, respectively, as shown in Fig.~\ref{fig:feq}. Thus, the sum rate at the proposed frequency (i.e., $300$~GHz) is approximately $40\%$ higher than at $28$~GHz and $87\%$ higher than at $6$~GHz.}

 \subsubsection{Impact of the Network Area}
\begin{figure}[!htp] 
 \centering \includegraphics[width=0.6\linewidth]{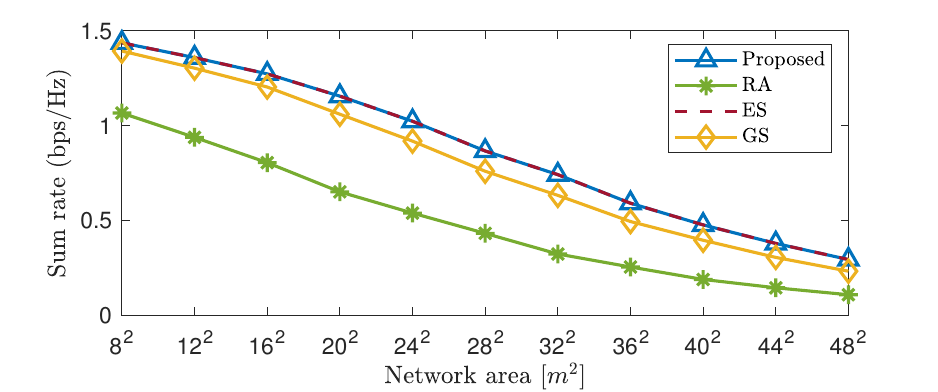}  \caption{{Sum rate versus network area, for $N=100\times100$, $K=64$, $I=J=10$, and $P=23$ dBm.}} \label{R_area} 
 \end{figure}

Fig.~\ref{R_area} shows that the sum rate decreases with increases in the network size for all schemes due to increased distance between devices. In particular, the pathloss is distance-dependent, so as distance increases, the sum rate decreases. In addition, the sum rate achieved by the proposed scheme is approximately identical to that of the ES scheme and superior to the sum rates of the GS and RA schemes, as shown in Fig.~\ref{R_area}.
\subsubsection{Impact of the Mobility}
 \begin{figure}[!htp] 
 \centering \includegraphics[width=0.5\linewidth]{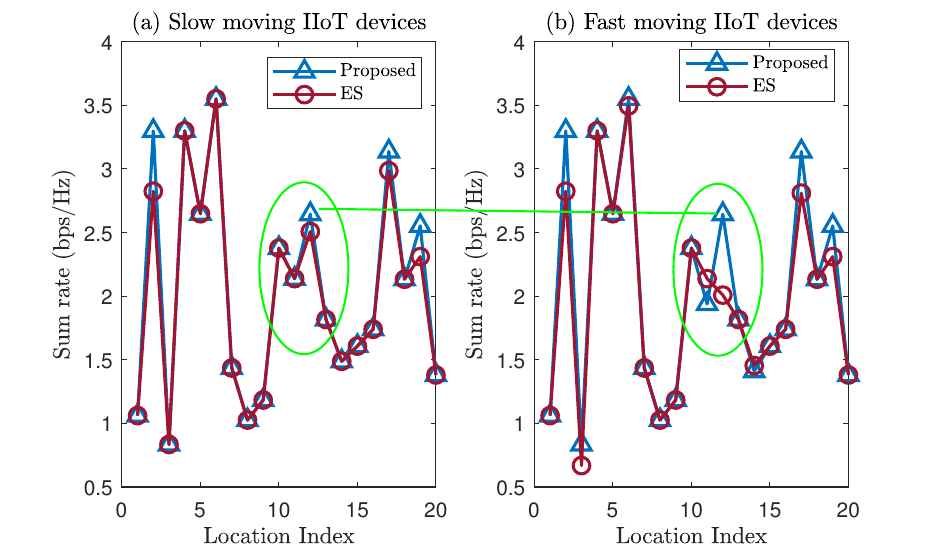}  \caption{Sum rate versus location index for different mobility scenarios, for a network area of $40\times40$ $\mathrm{m}^2$, $K=64$, and $P=23$ dBm.} \label{R_sf} 
 \end{figure}
In Fig.~\ref{R_sf}, the performance of the proposed JIIA and the ES algorithms is compared in different mobility scenarios. The JIIA algorithm achieves a similar sum rate in both slow and fast mobility scenarios, indicating that it is robust to changes in mobility. In contrast, the ES algorithm performs comparably to the JIIA algorithm only in scenarios with slow mobility. However, in scenarios with fast mobility, the ES algorithm achieves a lower sum rate than the proposed JIIA algorithm. This is because the ES algorithm requires a longer period to search for the optimal association, while the pause time is short in fast mobility scenarios.

\subsubsection{{Impact of the CSI}}
\begin{figure}[!htp] 
 \centering \includegraphics[width=0.6\linewidth]{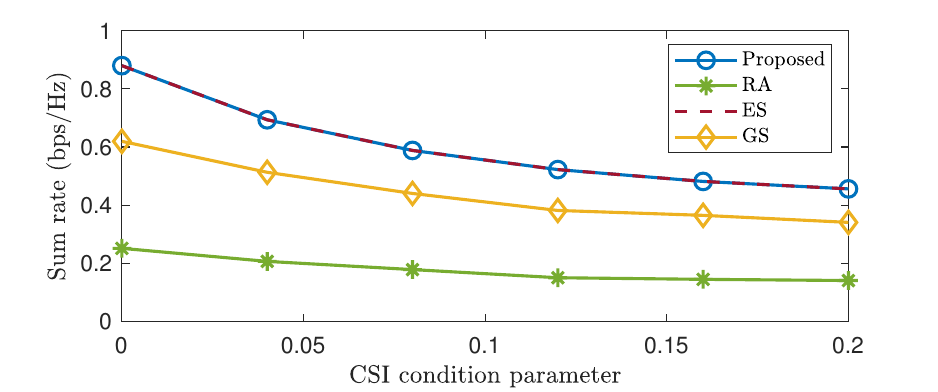} \caption{{Sum rate versus CSI condition parameter, for a network area of $40\times40$ $\mathrm{m}^2$, $N=100\times100$, $I=J=10$, and $K=64$.}} \label{imperfect_CSI1}
 \end{figure}

{
In this section, we demonstrate the impact imperfect CSI has and compare the results obtained with perfect CSI and imperfect CSI. Fig.~\ref{imperfect_CSI} shows that the sum rate is lower with imperfect CSI for all the schemes. This is due to channel uncertainty. Furthermore, it can be observed that the proposed scheme performs similarly to the ES method in both the perfect and imperfect CSI cases. 
\begin{figure}[!htp] 
 \centering \includegraphics[width=0.6\linewidth]{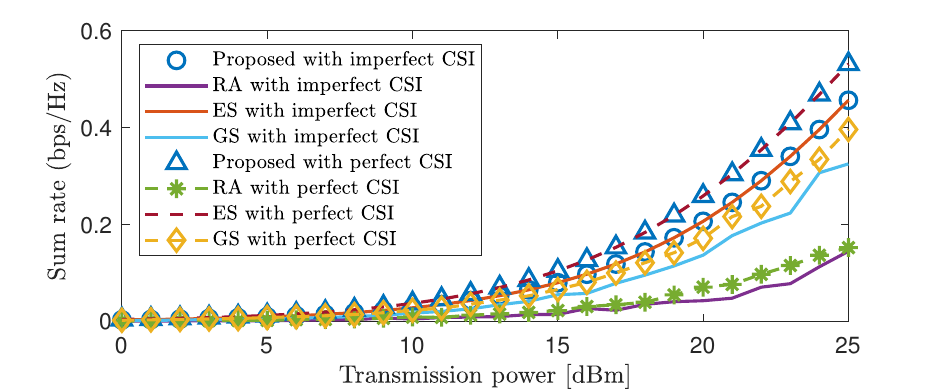} \caption{{Sum rate versus transmission power, for a network area of $40\times40$ $\mathrm{m}^2$, $N=100\times100$, $I=J=10$, and $K=64$.}} \label{imperfect_CSI}
 \end{figure}
In Fig.~\ref{imperfect_CSI1}, the sum rate is plotted against the CSI condition parameter. The results indicate the sum rate decreases with the CSI condition parameter for all the schemes. This is due to changes in the channel condition. Moreover, the results show that the sum rate of the proposed scheme is approximately similar to that of the ES method. Furthermore, the proposed scheme achieves a higher sum rate than the GS and RA schemes.} 
 
\subsection{Performance Comparison with Association Overhead}
The objective of this section is to assess the effectiveness of the proposed algorithm in terms of network sum rate while taking into account association overhead. {To estimate association overhead, we count the number of time slots required for JIIA and then divide it by the coherence interval. Furthermore, by subtracting this overhead duration from 1 and multiplying that number by the achievable rate, we obtain the achievable rate~\eqref{rate} with association overhead}. 
  \begin{figure}[!htp] 
 \centering \includegraphics[width=0.6\linewidth]{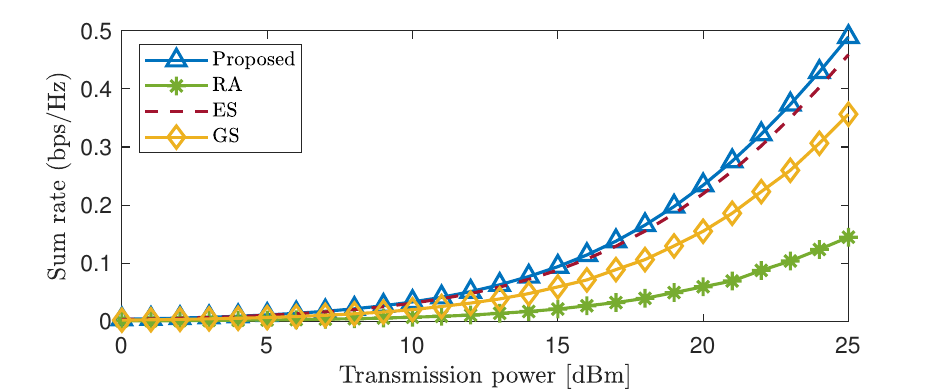} \caption{{Sum rate versus transmission power considering association overhead, for a network area of $40\times40$ $\mathrm{m}^2$, $N=100\times100$, $I=J=10$, and $K=64$.}} \label{R_Power_H}
 \end{figure}
The results presented in Fig.~\ref{R_Power_H} demonstrate the relationship between transmission power and sum rate, as well as the comparative performance of different algorithms. Across all considered cases, we observe that increasing the transmission power results in a corresponding increase in the sum rate. Moreover, the proposed algorithm achieves superior performance compared to the ES, GS, and RA schemes. This performance advantage is attributed to the proposed algorithm's ability to minimize the negative impact of the association overhead, which is a significant drawback of the ES scheme.
\section{Conclusions}\label{conc}
This paper investigated IRS-assisted THz communications to enable future smart industries for the IIoT, where several IRSs are deployed to assist with communication. {First, we formulated the channel model with perfect CSI and imperfect CSI. Then, we formulated an optimization problem to maximize the network sum rates. It was NP-hard, so we decomposed it to solve it.} We solved the PS configuration with a diffuse reflection model. Then, the decoding and beamforming vectors are achieved using the MMSE method. In this work, we aimed to solve the power allocation problem $(\vec{P3})$ and the JIIA problem $(\vec{P5})$. We achieved the analytical expression for the optimal power allocation for $(\vec{P3})$. Furthermore, we solved the JIIA problem $(\vec{P5})$ with our proposed matching-based JIIA algorithm. Simulation results validated the potential of the proposed IRS-assisted THz communication scheme and demonstrated its superiority over GS and RA schemes. We also analyzed the performance while considering the association overhead, and the results showed that the proposed scheme outperforms the ES scheme. This is because the ES scheme incurs a high association overhead, whereas our proposed scheme does not. Additionally, the computational complexity analysis confirmed that the proposed JIIA algorithm is more computationally efficient than the ES.

\appendices
\section{Proof of Theorem~\ref{T_NP}}\label{T_NP_proof}

To demonstrate that problem (9) is NP-hard using computational complexity theory, we can follow these three steps: (1) Select a well-known decision problem $Q$ that is already proven to be NP-complete. (2) Create a polynomial time conversion from any instance of $Q$ to an instance of problem~\eqref{eq_opt_prob}. (3) Prove that both instances have the same objective value after the transformation. We first assume the case of power allocation and JIIA with the given remaining variables. The sum-rate maximization problem in~\eqref{eq_opt_prob} can be expressed as
\begin{subequations}\label{eq_opt_prob_proof1}
\begin{alignat}{2}
& \underset{\vec{P},\vec{\Psi} }{\text{maximize}}
&\quad 
& \textstyle \sum_{l,m} R_{l,m} \Psi_{l,m}, \label{eq_optProbb}\\
&
\text{subject to} 
&& {\textstyle\sum_{j=1}^{J}p_j\Vert\overline{w}_{j}\Vert^2} \leq  \vec{P}_{\AP},\quad
\textstyle  p_{j}\geq 0  , \forall{j}, \label{eq_constraint3bb}\\
&&& \textstyle \sum_{l\in L} \Psi_{l,m}\leq 1 , \forall{l\in L}, \quad \textstyle \sum_{m\in M} \Psi_{l,m} \leq 1  , \forall{m\in M}, \label{eq_constraint55}\\
&&& \textstyle  \Psi_{l,m} \in \{0,1\}, \forall{l,m}, \label{eq_constraint66}
\end{alignat}
\end{subequations}
which has been proved to be NP-hard in~\cite{cui2017optimal}. Furthermore, we prove that the IRS association is NP-hard for the given power allocation matrix. The sum-rate maximization problem in~\eqref{eq_opt_prob} can be expressed as
\begin{alignat}{2}
& \underset{\vec{\Psi} }{\text{maximize}}
&\quad
& \textstyle \sum_{l,m} R_{l,m} \Psi_{l,m}, 
\quad\text{subject to} \quad \eqref{eq_constraint55},\eqref{eq_constraint66},
\end{alignat}
which is similar to the link scheduling
problem, a known NP-hard problem~\cite{mlika2018user}.
This completes the proof of Theorem \ref{T_NP}.
\section{Proof of Theorem~\ref{T_P}} \label{T_P_proof}
{The Lagrangian function associated with problem $(\vec{P3})$ in~\eqref{eq_opt_P3} can be expressed as
\begin{align} \label{L1}
    L(p_j,\mu)=\sum\limits_{\substack{j=1}}^J\log_2\bigg(1+\frac{p_j\left|\widehat{h}_{j,j}(t)\right|^2}{\textstyle\sum_{{j'=1, j'\neq j}}^J p_{j'} \left|h_{j,j'}(t) \right|^2+p_j\mathcal{X} +\sigma^2_j}\bigg)-\mu \big(\textstyle\sum_{{j=1}}^J p_j\Vert\overline{w}_{j}\Vert^2 - P_{\AP}\big)-\upsilon_j p_j,
\end{align}
where $\mathcal{X} = \sigma_{\widetilde{\vec{g}}_{j, m}}^2    \Vert \widehat{\vec{G}}_m \overline{\vec{w}}_{j} \Vert^2
+ ( \sigma_{\widetilde{\vec{g}}_{j, m}}^2\sigma_{\widetilde{\vec{G}}_{m}}^2 +\sigma_{{\widetilde{\vec{G}}_m}}^2 \Vert \widehat{\vec{g}}_{j, m} \Vert^2)  \Vert \overline{\vec{w}}_{j} \Vert^2$.
By setting the derivative of~\eqref{L1} with respect to $p_j$ to zero, we can obtain the KKT system of the optimization problem as
\begin{align} \label{L2}  
    & \frac{\partial L(p_j,\mu)}{\partial p_j} = 0\nonumber \\
    & p_j\Big(1+\frac{\mathcal{X} }{\left|\widehat{h}_{j,j}(t)\right|^2}\Big)+\frac{\sum_{{j'=1, j'\neq j}}^J p_{j'} \left|h_{j,j'}(t) \right|^2 +\sigma^2_j}{\left|\widehat{h}_{j,j}(t)\right|^2}=\frac{1}{\mu  \Vert\overline{w}_{j}\Vert^2 +\upsilon_j},
\end{align}
where $\mu$ and $\upsilon_j$ are the Lagrangian multipliers. The multiplier $\upsilon_j$ stands for the effect of interference caused by the $j$-th device on other devices and is called the taxation term~\cite{lee2008simplified}. The interference term can be represented as
\begin{align}\label{int}
    \iota_j(t)&= \textstyle\sum_{{j'=1, j'\neq j}}^J p_{j'} \left|h_{j,j'}(t) \right|^2,
    \nonumber\\&=\sum\limits_{j'=1, j' \ne j}^J\Big( p_{j'} \left| {(\widehat{\vec{h}}_j^{\DIRS_m}(t))^{\sf H} \overline{\vec{w}}_{j'}}\right|^2 
+  \big[
    \sigma_{\widetilde{\vec{g}}_{j, m}}^2
    \big\Vert \widehat{\vec{G}}_m \overline{\vec{w}}_{j'} \big\Vert^2
+   \big( \sigma_{\widetilde{\vec{g}}_{j, m}}^2
        \sigma_{\widetilde{\vec{G}}_{m}}^2 
        + \sigma_{{\widetilde{\vec{G}}_m}}^2 \Vert \widehat{\vec{g}}_{j, m} \Vert^2
    \big)  \Vert \overline{\vec{w}}_{j'} \Vert^2
\big]\Big). 
\end{align}
To find the power of the $j$-th device, we rearrange~\eqref{L2} and utilize~\eqref{int} as
\begin{align} 
&{p_j\Big(1+\frac{\mathcal{X} }{\left|\widehat{h}_{j,j}(t)\right|^2}\Big)+\frac{\textstyle\sum_{{j'=1, j'\neq j}}^J p_{j'} |h_{j,j'}(t) |^2}{|\widehat{h}_{j,j}(t) |^2}+\frac{\sigma^2_j}{|\widehat{h}_{j,j}(t) |^2}}=\frac{1}{\mu \Vert\overline{w}_{j}\Vert^2 +\upsilon_j}\nonumber\\
&p_j\Big(1+\frac{\mathcal{X} }{|\widehat{h}_{j,j}(t)|^2}\Big)=\frac{1}{\mu \Vert\overline{w}_{j}\Vert^2 +\upsilon_j}-\frac{\iota_j(t)+\sigma^2_j}{|\widehat{h}_{j,j}(t) |^2}\nonumber
\end{align}
\begin{align} \label{L3} 
&p_j=\left[\frac{\frac{1}{\mu \Vert\overline{w}_{j}\Vert^2+\upsilon_j}-\frac{ \iota_j(t)+\sigma^2_j}{\left|\widehat{h}_{j,j}(t) \right|^2}}{1+\frac{\mathcal{X} }{\left|\widehat{h}_{j,j}(t)\right|^2}}\right]^{+}=\left[\frac{\frac{1}{\mu \Vert\overline{w}_{j}\Vert^2+\upsilon_j}-\frac{ \iota_j(t)+\sigma^2_j}{\left|\widehat{h}_{j,j}(t) \right|^2}}{1+\frac{\sigma_{\widetilde{\vec{g}}_{j, m}}^2    \Vert \widehat{\vec{G}}_m \overline{\vec{w}}_{j} \Vert^2
+ ( \sigma_{\widetilde{\vec{g}}_{j, m}}^2\sigma_{\widetilde{\vec{G}}_{m}}^2 +\sigma_{{\widetilde{\vec{G}}_m}}^2 \Vert \widehat{\vec{g}}_{j, m} \Vert^2)  \Vert \overline{\vec{w}}_{j} \Vert^2}{\left|\widehat{h}_{j,j}(t)\right|^2}}\right]^{+},
\end{align}
where $[x]^+ = \max \{x,0\}$. Given $\iota_{j}$ and $\upsilon_j$, the water-filling algorithm can find the optimal $p_j$. Substituting~\eqref{L3} into constraint~\eqref{eq_constraint_P3_1} yields
\begin{align} \label{L4} 
\vec{P}_{\AP}=\sum_{j=1}^{J}\left[{\Vert\overline{w}_{j}\Vert^2}\Bigg(\frac{\frac{1}{\mu \Vert\overline{w}_{j}\Vert^2+\upsilon_j}-\frac{ \iota_j(t)+\sigma^2_j}{\left|\widehat{h}_{j,j}(t) \right|^2}}{1+\frac{\sigma_{\widetilde{\vec{g}}_{j, m}}^2    \Vert \widehat{\vec{G}}_m \overline{\vec{w}}_{j} \Vert^2
+ ( \sigma_{\widetilde{\vec{g}}_{j, m}}^2\sigma_{\widetilde{\vec{G}}_{m}}^2 +\sigma_{{\widetilde{\vec{G}}_m}}^2 \Vert \widehat{\vec{g}}_{j, m} \Vert^2)  \Vert \overline{\vec{w}}_{j} \Vert^2}{\left|\widehat{h}_{j,j}(t)\right|^2}}\Bigg)\right]^{+} .
\end{align}

This completes the proof of Theorem \ref{T_P}.}
\section{Proof of Theorem~\ref{T1}} \label{T1_proof}
{The MINLP problem formulated is NP-hard since it combines optimizing over discrete variables with nonlinear functions. In this context, the ES scheme is conventionally used to find the optimal solution~\cite{burer2012non}. The optimality of the ES method can be proven by the following conditions.
\begin{itemize}
    \item The ES method's characteristics dictate that, at every iteration, the sum rate for the current association matrix is calculated as in step 7 of Algorithm~\ref{algo_es}.
    \item Then, in steps 9 to 12 of Algorithm~\ref{algo_es}, we check if the sum rate of the current association matrix is greater than that of the previous one and select the matrix with the higher sum rate.
\end{itemize}
This way, we obtain the optimal association matrix $\vec{\Psi}^\star$, resulting in the highest sum rate. 

Now, we prove that the JIIA matrix achieved with the proposed algorithm is stable and optimal.
First, we note that the algorithm must terminate in polynomial time, as indicated in Remark~\ref{R2}. When it terminates, we have a match. Let's prove it is stable and optimal. Suppose the proposed algorithm matches $\UIRS_l$ and $\DIRS_m$. However, $\UIRS_l$ prefers $\DIRS_{m'}$ over $\DIRS_{m}$. When the algorithm is run, $\UIRS_l$ offers to $\DIRS$s in order of preference, so if $\UIRS_l$ ends up with $\DIRS_m$, it must have proposed to $\DIRS_{m'}$ at some point and been rejected. Now, when $\DIRS_{m'}$  rejected $\UIRS_{l}$, it must have had another offer in hand that it preferred over that of $\UIRS_{l}$. As the algorithm proceeded, $\DIRS_{m'}$  must have received a better offer, so it ended up with a $\UIRS$ that is preferred over $\UIRS_l$. Therefore, $\DIRS_{m'}$  will not form a blocking pair with $\UIRS_{l}$. It follows that, at the end of the algorithm, there will be no better uplink IRS $\UIRS_{l}$  that can form a blocking pair with downlink IRS $\DIRS_{m}$ than its current partner. Hence the matching is stable and optimal. This completes the proof of Theorem \ref{T1}.}

\bibliographystyle{IEEEtran}
\bibliography{main}
\end{document}